\documentclass[aps,prd,superscriptaddress,nofootinbib,eqsecnum,twocolumn]{revtex4}

\pdfoutput=1

\usepackage{amsfonts}
\usepackage{amsmath}
\usepackage{amssymb}
\usepackage{graphicx,color}
\usepackage{float}
\usepackage{hyperref}
\usepackage{subfigure}
\usepackage{mathtools,slashed}
\usepackage{ulem}
\usepackage{xcolor}

\def\be {\begin{equation}}
\def\ee {\end{equation}}
\def\bea {\begin{eqnarray}}
\def\eea {\end{eqnarray}}
\def\bc {\begin{center}}
\def\ec {\end{center}}

\def\nn {\nonumber}

\date{\today}

\begin{document}

\title{Hot QCD at finite isospin density: confronting SU(3) Nambu-Jona-Lasinio model with recent lattice data}

\author{Bruno. S. Lopes}
\affiliation{Departamento de F\'{i}sica, Universidade Federal de Santa Maria, Santa Maria, RS 97105-900, Brazil}

\author{Sidney S. Avancini}
\affiliation{Departamento de F\'{i}sica, Universidade Federal de Santa Catarina, 88040-900 Florian\'{o}polis, 
Santa Catarina, Brazil}

\author{Aritra Bandyopadhyay}
 \affiliation{Guangdong Provincial Key Laboratory of Nuclear Science, Institute of Quantum Matter, South China Normal University, Guangzhou 510006, China}
 \affiliation{Guangdong-Hong Kong Joint Laboratory of Quantum Matter, Southern Nuclear Science Computing Center, South China Normal University, Guangzhou 510006, China}

\author{Dyana C. Duarte}
\affiliation{Institute for Nuclear Theory, University of Washington, Seattle, WA 98195, USA}
\affiliation{Departamento de F\'{i}sica, Instituto Tecnol\'ogico de 
Aeron\'autica, 12228-900 S\~ao Jos\'e dos Campos, SP, Brazil}

\author{Ricardo L. S. Farias}
\affiliation{Departamento de F\'{i}sica, Universidade Federal de Santa Maria, Santa Maria, RS 97105-900, Brazil}

\begin{abstract}

Extending our recently published $SU(2)$ results for zero temperature we now compute the QCD equation of state for finite isospin density within the three flavor Nambu-Jona-Lasinio model in the mean field approximation, motivated by the recently obtained Lattice QCD results for both zero and finite temperatures. Like our previous study, here also we have considered both the commonly used Traditional cutoff Regularization Scheme and the
Medium Separation Scheme. Our results are compared with recent high-precision lattice simulations as well as previously obtained results in two-flavor Nambu-Jona-Lasinio model. The agreement between the lattice results and the predictions from three-flavor NJL model is very good for low values of $\mu_I$ (for both zero and finite temperatures). For larger values of $\mu_I$, the agreement between lattice data and the two-flavor NJL predictions is surprisingly good and better than with the three-flavor predictions.
 
\end{abstract}

\maketitle


\section{Introduction}
\label{sec1}

As the fundamental theory of strong interactions, the phase structure of Quantum Chromodynamics (QCD) has been studied from different angles over the years. Aside from the well explored systems at finite temperatures and finite baryon densities, several new dimensions have been added to the QCD phase diagrams: isospin chemical potential, magnetic field, electric field, rotation to name a few. Though near future relativistic Heavy-Ion-Collision (HIC) experiments in FAIR and NICA have been continuing to inspire studies of physical systems at low temperatures and finite baryon densities such as neutron stars~\cite{Fukushima:2010bq,Alford:2007xm}, theoretical hurdles are still there, predominantly due to the well known fermion ``sign problem''~\cite{karsch,Muroya:2003qs} encountered by non-perturbative lattice calculations. Lattice QCD's recent progress with the sign problem can be monitored in Ref~\cite{Bedaque:2017epw}. 

Among the relatively new features of the QCD phase diagram, finite isospin chemical potential ($\mu_I$) plays an important role, specially because unlike finite baryon chemical potential it does not suffer from the sign problem in lattice QCD based calculations. First bunch of lattice QCD results at finite temperature and isospin density
appeared in early 2000's~\cite{Kogut:2002zg,Kogut:2002tm} with dynamical $u$ and $d$ quarks, although with unphysical pion masses and/or an unphysical flavour content. This followed various studies by other available theoretical tools yielding qualitatively similar results. These studies include chiral perturbation theory ($\chi$PT)~\cite{Son:2000xc,Son:2000by,Splittorff:2000mm,Loewe:2002tw,Loewe:2005yn,Fraga:2008be,Cohen:2015soa,Janssen:2015lda,Carignano:2016lxe,Lepori:2019vec,Adhikari:2019mdk,Adhikari:2019mlf,Adhikari:2020ufo}, Hard Thermal Loop perturbation theory (HTLPt)~\cite{Andersen:2015eoa}, Nambu-Jona-Lasinio (NJL) model~\cite{Frank:2003ve,Toublan:2003tt,Barducci:2004tt,He:2005sp,He:2005nk,He:2006tn,Ebert:2005cs,Ebert:2005wr,Sun:2007fc,Andersen:2007qv,Abuki:2008wm,Mu:2010zz,Xia:2013caa,Khunjua:2018jmn,Khunjua:2018sro,Khunjua:2017khh,Khunjua:2019lbv,Khunjua:2019ini,Khunjua:2020xws,Ebert:2016hkd,Lu:2019diy} and its Polyakov loop extended version PNJL~\cite{Mukherjee:2006hq,Bhattacharyya:2012up}, quark meson model (QMM)~\cite{Kamikado:2012bt,Ueda:2013sia,Stiele:2013pma,Adhikari:2018cea}. Recently, early lattice QCD results have been modified by using an improved lattice action with staggered fermions at physical quark masses and results for finite isospin density are presented in Refs~\cite{Brandt:2016zdy,Brandt:2017zck,Brandt:2017oyy,Brandt:2018wkp}.


In our recent work along the similar line~\cite{Avancini:2019ego}, we focused on a new type of compact stars known as pion stars~\cite{Carignano:2016lxe,Brandt:2018bwq}, where the pion condensates are considered to be the dominant constituents of the core under the circumstance of vanishing neutron density. In spite of being a subset of boson stars~\cite{Wheeler:1955zz,Kaup:1968zz,Jetzer:1991jr,Colpi:1986ye,Liebling:2012fv}, pion stars are free from hypothetical beyond standard model contributions like QCD axion. This gave us a scenario to work with finite isospin density along with zero temperature and zero baryon density which bypasses the sign problem unlike systems with high baryon densities. Hence it is easily accessible through first principle methods~\cite{Brandt:2018bwq} and through the pion stars' Equation of State (EoS) we now know about its large mass and radius in
comparison with neutron stars~\cite{Brandt:2018bwq,Andersen:2018nzq}. Our study in the said context of pion stars within two flavor NJL model showed better quantitative agreement with the lattice QCD results than similar studies within the chiral perturbation theory~\cite{Adhikari:2019mdk}. 

Unlike our last study~\cite{Avancini:2019ego} where we have only considered the setting of pion stars with zero temperature, in the present work we plan also to explore the systems with finite temperature. Early universe with very high temperature has been known to have possibilities of pion condensation driven by high lepton asymmetry~\cite{Abuki:2009hx,Schwarz:2009ii,Wygas:2018otj}. Furthermore, in this work we have extended our two-flavor study within the three flavor NJL model. While the two-flavor studies are 
sufficient to describe the pion condensation at finite isospin density, a three-flavor study gives us the chance to explore the roles of the strangeness degree of freedom and the $U_A(1)$ anomaly in the present context. Hence in our three flavor NJL Lagrangian, we will also be considering the Kobayashi-Maskawa-'t Hooft (KMT) term, which mimics the $U_A(1)$ anomaly in the NJL model.

Just like QCD systems with $\mu_I \neq 0,~\mu_B=\mu_s=T=0$~\cite{Brandt:2018bwq}, QCD systems with $\mu_I \neq 0,~\mu_B=0,~\mu_s\neq 0,~T\neq 0$ are also being explored well within lattice QCD~\cite{Brandt:2017oyy,Brandt:2018wkp,Brandt:2019idk,Brandt:pc,Vovchenko:2020crk}. Successful premises of this work have already been provided by our previous study~\cite{Avancini:2019ego} which showed an exceptional quantitative agreement between NJL and lattice QCD results. On the basis of that and recent improved three flavor lattice QCD results at zero and finite temperature~\cite{Brandt:2017oyy,Brandt:2018wkp,Brandt:2018bwq,Brandt:2019idk,Brandt:pc,Vovchenko:2020crk} give us the perfect opportunity for the consistency check of the NJL model. 
As in our previous work~\cite{Avancini:2019ego} we have tried to rectify the regularization issues within NJL model to deal with the cutting of important degrees of freedom near the Fermi surface because of a sharp ultraviolet (UV) cutoff~\cite{Farias:2005cr,Farias:2006cs,Braguta:2016aov}. Besides the commonly used Traditional Regularization Scheme (TRS) we have used the Medium Separation Scheme (MSS)~\cite{Farias:2005cr,Farias:2016let,Duarte:2018kfd}, which properly separates the medium effects from divergent integrals. For systems with high values of $\mu_I$ ($\sim\Lambda$) the role of MSS becomes more and more important. 

The paper is organized as follows. In section~\ref{sec2} we discuss the basic formalism of
the three-flavor NJL model both within TRS and MSS. In section \ref{sec3} we present and discuss our results obtained with the traditional regularization scheme and with the medium separation scheme, for both zero and finite temperature. Thermodynamic results and the $T-\mu_I$ phase diagram are also presented and compared with other state of the art calculations.

\section{Formalism}
\label{sec2}

We start with the partition function for the three-flavor NJL model at finite baryonic and isospin chemical potential, given by 
\bea
&&Z_{\textrm{NJL}} (T,\mu_B,\mu_I,\mu_S) = \int[d\bar{\psi}][d\psi]\times \nn\\
&&\exp\left[\int\limits_0^\beta d\tau \int d^3x \left(\mathcal{L}_{\textrm{NJL}} +\bar{\psi} \hat{\mu} \gamma_0 \psi \right)\right], 
\label{partition_f}
\eea
where the quark chemical potential matrix in flavor space is
\bea
\hat{\mu} = \begin{pmatrix}
             \mu_u ~~~~ 0 ~~~~0 \\ 0 ~~~~ \mu_d ~~~~ 0 \\ 0~~~~0~~~~\mu_s
            \end{pmatrix},
\eea
and $\mu_{u,d,s}$ can be expressed in terms of the baryonic, the isospin and the strangeness chemical potential as 
\bea
\mu_u &=& \frac{\mu_B}{3} + \mu_I, \nn\\
\mu_d &=& \frac{\mu_B}{3} - \mu_I, \nn \\
\mu_s &=& \frac{\mu_B}{3} -\mu_S, \nn
\eea
such that $\mu_I = (\mu_u-\mu_d)/2$.
$\mathcal{L}_{\textrm{NJL}}$ appearing in Eq.~\eqref{partition_f} is the NJL Lagrangian considering scalar
and pseudoscalar interactions, i.e. 
\bea
\mathcal{L}_{\textrm{NJL}} &=& \bar{\psi}\left(i\slashed{\partial}-m\right)\psi + G \sum_{\alpha=0}^{N_f^2-1} \left[\left(\bar{\psi}\lambda_\alpha\psi\right)^2 +\left(\bar{\psi}i\gamma_5\lambda_\alpha\psi\right)^2\right]\nn\\
&& - K \left[\textrm{det}~ \bar{\psi} (1+\gamma_5) \psi + \textrm{det}~ \bar{\psi} (1-\gamma_5) \psi\right] \nn
\label{lag_njl}
\eea
where $\psi = (u ~d~s)^T$ and $m = \textrm{diag}(m_u,m_d,m_s)$ represent the quark fields and their current mass respectively and $G$ is the scalar 
coupling constant of the model from the four-fermion interaction. The last term is the KMT term which represents the breaking of the flavor symmetry in the chiral limit due to $U_A(1)$ anomaly. $K$ is also known as the $U_A(1)$ breaking strength. 

Next, in the mean field approximation we introduce the chiral condensates 
\bea
\sigma_l = \sigma_{u/d} = -4G\langle \bar{u}u\rangle / \langle \bar{d}d\rangle, ~~ \sigma_s = -4G\langle \bar{s}s\rangle,
\eea
and the pseudoscalar pion condensate 
\bea
\Delta = 2(2G-K\sigma_s)\langle \bar{u}i\gamma_5d\rangle,
\eea
where $\Delta$ can be considered as real without loss of generality as the related phase factor can be arbitrarily chosen due to the spontaneously broken $U_I(1)$ symmetry\footnote{It is important to note here that in this work we are not considering kaon condensation as we are working in the limit of vanishing baryonic and strangeness chemical potentials, i.e. $\mu_B=\mu_S=0$.}~\cite{Xia:2013caa}. In terms of these condensates, the thermodynamic potential for $N_f=2+1$ in the mean field approximation is given as 
\bea
&&\Omega(\sigma_l,\sigma_s,\Delta) = \frac{2\sigma_l^2+\sigma_s^2}{8G}+\frac{K\sigma_s\sigma_l^2}{16G^3} +\left(G+\frac{K\sigma_s}{4G}\right)\times\nn\\
&&~~\frac{\Delta^2}{\left(2G+\frac{K\sigma_s}{4G}\right)^2}-2N_c\int\limits_k^\Lambda\Bigg[E_k^++E_k^-+E_k^s+2T \times \nn\\
&&~~\ln\left\{ \left(1+e^{-\beta E_k^-}\right)\left(1+e^{-\beta E_k^+}\right)\left(1+e^{-\beta E_k^s}\right)\right\}\Bigg]
\label{EffPot}
\eea
with $E_k^\pm=\sqrt{\left(E_k^l\pm\frac{\mu_I}{2}\right)^2+\Delta^2}$ with $E_k^l=\sqrt{k^2+M_l^2}$ and $E_k^s=\sqrt{k^2+M_s^2}$ and the symbol $\int_k^{\Lambda}$ indicates three momentum integrals that need to be regularized. Different effective masses $M_l$ and $M_s$ are defined as 
\bea
M_l &=& m_l + \sigma_l + \frac{K\sigma_l\sigma_s}{8G^2}, \nn\\
M_s &=& m_s + \sigma_s + \frac{K\sigma_l^2}{8G^2} + \frac{K}{2}\frac{\Delta^2}{\left(2G+\frac{K\sigma_s}{4G}\right)^2}, \nn
\eea
where $m_l = m_u =m_d$ and $m_s$ are the current quark masses. 

The ground state at finite temperature and isospin chemical potential is determined by minimizing $\Omega(\sigma_l,\sigma_s,\Delta)$ with respect to $\sigma_l$, $\sigma_s$ and $\Delta$, i.e. by solving $\partial\Omega / \partial \sigma_l = \partial\Omega / \partial \sigma_s = \partial\Omega / \partial \Delta = 0$. 

In the following subsections we discuss in more details different ways
of regularizing these integrals.
 The thermodynamic quantities, i.e. the pressure, the isospin density and the energy density 
 of the system are then respectively given by
\bea
P_{\textrm{NJL}} &=& - \Omega_{\textrm{NJL}}(\sigma_{l/s}=\sigma_{l/s}^0; \Delta = \Delta^0),\label{P_njl}\\
\langle n_I\rangle_{\textrm{NJL}} &=& \frac{\partial P_{\textrm{NJL}}}{\partial \mu_I},\label{nI_njl} \\ 
\varepsilon_{\textrm{NJL}} &=& -P_{\textrm{NJL}} + \mu_I \langle n_I\rangle_{\textrm{NJL}}+T\frac{\partial P_{\textrm{NJL}}}{\partial T}. \label{ed_njl}
\eea
Finally, the interaction measure (or trace anomaly) within the NJL model is given by the relation between $P_{\textrm{NJL}}$ and $\varepsilon_{\textrm{NJL}}$, i.e.,
\be
I_{\textrm{NJL}} = \varepsilon_{\textrm{NJL}} - 3P_{\textrm{NJL}}.
\label{tA_njl}
\ee


\subsection{Regularization}

Due to the nonrenormalizable nature of the NJL model, any physical quantity determined in its framework 
will be dependent on the scale of the model $\Lambda$. In the SU(2) version, the usual regularization schemes 
consist in to determine $\Lambda$, the coupling constant 
$G$ and current quark mass $m_u = m_d = m_c$ that reproduce the empirical values of the pion mass $m_{\pi}$, 
the pion decay constant $f_{\pi}$ and the quark condensate $\left\langle\bar{\psi}\psi\right\rangle$. 
Since our aim is to compare our results with lattice simulations we have used two different sets of parameters,
for $T = 0$ and $T\neq 0$, as can be seen in Tab.~\ref{parameter_setsSU2}.

\begin{table}
\caption{Different parameter sets are listed, which have been used in the present study for SU(2) case. }
\begin{center}
\begin{tabular}{c|c|c}
\hline Sets & Input parameters & Output parameters \\ \hline 
   & $f_\pi = 93$ MeV                                & $\Lambda = 659.325$ MeV \\ 
I  & $m_\pi = 131.7$ MeV                             & $G = 2.07835/\Lambda^2$ \\ 
   & $\langle\bar{\psi}\psi\rangle^{1/3}$ = 250 MeV  & $m_c=4.757$ MeV   \\ \hline 
   & $f_\pi = 92.4$ MeV                              & $\Lambda = 659.325$ MeV\\  
II & $m_\pi = 135.5263$ MeV                          & $G =2.07691/\Lambda^2$ \\ 
   & $\langle\bar{\psi}\psi\rangle^{1/3}$ = 250 MeV  & $m_c=4.93651$ MeV \\ \hline 
\end{tabular}
\end{center}
\label{parameter_setsSU2}
\end{table}

The SU(3) case is much more complicated; the procedure is shown in details in appendix-\ref{ap_su3} and 
the values obtained are shown in Tab.~\ref{parameter_setsSU3}.
The two sets represent the parameters we use for two different cases, 
set-I for the case of zero temperature and set-II for the case of finite temperature, following the value 
of $m_\pi$ used by Lattice QCD respectively for both the cases. 

\begin{table}
\caption{Different parameter sets are listed, which have been used in the present study for SU(3) case.}
\begin{center}
\begin{tabular}{c|c|c}
\hline Sets & Input parameters & Output parameters \\ \hline 
   & $f_\pi = 93$ MeV          & $\Lambda = 574.68$ MeV \\ 
   & $m_\pi = 131.7$ MeV       & $G = 2.2066/\Lambda^2$ \\ 
I  & $m_K=490$ MeV             & $K = 10.426/\Lambda^5$ \\ 
   & $m_\eta =950$ MeV         & $m_s = 140$ MeV \\ 
   & $m_l=5.3$ MeV             &  \\ \hline 
   & $f_\pi = 92.4$ MeV        & $\Lambda = 608.431$ MeV\\  
   &  $m_\pi = 135.5263$ MeV   & $G =1.782/\Lambda^2$ \\ 
II & $m_K=497.7$ MeV           & $K = 12.8525/\Lambda^5$ \\ 
   & $m_\eta=957.8$ MeV        & $m_s= 140.305$ MeV \\ 
   & $m_l=5.5$ MeV             &  \\ \hline 
\end{tabular}
\end{center}
\label{parameter_setsSU3}
\end{table}

The fact that all physical quantities are dependent on $\Lambda$ does not mean that we can just naively use this 
cutoff in all integrals, since it may lead to incorrect results.  
In this work we compare the results of two different schemes, namely the {\it traditional re\-gu\-la\-ri\-za\-tion scheme} (TRS) and the {\it medium separation scheme} (MSS). The TRS is the most common in NJL studies, and consists only in to perform up to $\Lambda$
the integrals that do not depend on the temperature, e.g., the first three terms between brackets of the integral in 
Eq.~\eqref{EffPot} and its correspondents in the gap equations and thermodynamic quantities, while thermal integrals 
are performed up to infinity.

By the other hand, MSS provides a tool to disentangle all the medium dependencies from divergent 
contributions, so that only vacuum integrals remain to be regularized. This scheme has been applied 
to the NJL model and successfully shows qualitative agreement with lattice simulations and 
more elaborated theories, as might be seen in Refs.~\cite{Farias:2005cr,Farias:2016let,Duarte:2018kfd,Avancini:2019ego}.
Let us start for example from integral $I_{\Delta}$ of $\Delta$ gap equation:
\be
I_{\Delta} = \sum_{j= \pm 1}\int_{\Lambda} \frac{d^3k}{(2\pi)^3}
\frac{1}{\sqrt{(E_k^l + j\mu)^2 + \Delta^2}}\;,\label{Id}
\ee
whose TRS version is obtained just by making the replacement 
$\int_{\Lambda} \frac{d^3k}{(2\pi)^3} \to \int_0^{\Lambda} dk\frac{k^2}{2\pi^2}$. To use MSS we first 
rewrite 
\be
I_{\Delta} = \frac{1}{\pi}\sum_{j= \pm 1}\int\limits_{-\infty}^{+\infty}dx\int_{\Lambda} \frac{d^3k}{(2\pi)^3}
\frac{1}{x^2 + (E_k^l + j\mu)^2 + \Delta^2}\;,\label{Id4}
\ee
where, to ease the notation, we made the replacement $\mu_I/2\to\mu$. Using the identity
\bea
\lefteqn{ \frac{1}{x^2 + (E_k^l + j\mu)^2 + \Delta^2}} \nn\\
&& = \frac{1}{x^2 + k^2 + M_0^2}\nn\\
&& + \frac{M_0^2 - \Delta^2 - \mu^2 - M^2 - 2j\mu E_k^l}{(x^2 + k^2 + M_0^2)
\left[x^2 + (E_k^l + j\mu)^2 + \Delta^2\right]}.\;\;\;\label{ident}
\eea
Here $M_0$ is the vacuum mass of light quarks, obtained in the $T = \mu = \Delta = 0$ limit. After two iterations 
of this identity we obtain
\bea
\lefteqn{\sum_{j= \pm 1}\frac{1}{x^2 + (E_k^l + j\mu)^2 + \Delta^2}}\nn\\
&& = \frac{2}{x^2 + k^2 + M_0^2} + \frac{2\mathcal{M}}{(x^2 + k^2 + M_0^2)^2}\nn\\
&& + \frac{2\mathcal{M}^2 + 8\mu^2 (E_k^l)^2}{(x^2 + k^2 + M_0^2)^3}\nn\\
&& + \sum_{j= \pm 1}\frac{(\mathcal{M} - 2j\mu E_k^l)^3}{(x^2 + k^2 + M_0^2)^3
\left[x^2 + (E_k^l + j\mu)^2 + \Delta^2\right]}\;,\nn\\
\eea
where we have defined $\mathcal{M} = M_0^2 - \Delta^2 - \mu^2 - M_l^2$. After some manipulations 
and performing the integration in $x$ indicated in (\ref{Id4}) we obtain
\bea
\lefteqn{I_{\Delta}^{\rm MSS} = 2I_{\rm quad}(M_0)} \nn\\
&&- (M_l^2 - M_0^2 + \Delta^2 - 2\mu^2)I_{\rm log}(M_0)\nn\\
&&+ \left[\frac{3(\mathcal{M}^2 + 4\mu^2M_l^2)}{4} - 3\mu^2M_0^2\right]I_1 + 2I_2\;,
\eea
with the definitions
\begin{align}
&I_{\rm quad}(M_0) = \int\frac{d^3k}{(2\pi)^3}\frac{1}{\sqrt{k^2 + M_0^2}}\;,\\
&I_{\rm log}(M_0) = \int\frac{d^3k}{(2\pi)^3}\frac{1}{(k^2 + M_0^2)^{\frac{3}{2}}}\;,\\
&I_1 = \int\frac{d^3k}{(2\pi)^3}\frac{1}{(k^2 + M_0^2)^{\frac{5}{2}}}\;,\\
&I_2 = \frac{15}{32}\sum_{j= \pm 1}\int\frac{d^3k}{(2\pi)^3}\int\limits_0^1 dt (1-t)^2\nn\\
&\times\frac{(\mathcal{M} - 2j\mu E_k^l)^3}{\left[(2j\mu E_k^l - \mathcal{M})t + k^2 + M_0^2\right]^{\frac{7}{2}}}\;.
\label{i_2}
\end{align}

A similar procedure can be used to obtain the integrals of other quantities:
\bea
I_{\sigma_l} &=& \sum_{j= \pm 1}\int_{\Lambda} \frac{d^3k}{(2\pi)^3}\frac{1}{E_k^l}\frac{E_k^l + j\mu}{\sqrt{(E_k^l + j\mu)^2 + \Delta^2}}\;,\\
I_{n_I} & = & \sum_{j= \pm 1}\int_{\Lambda} \frac{d^3k}{(2\pi)^3}j\frac{E_k^l + j\mu}{\sqrt{(E_k^l + j\mu)^2 + \Delta^2}}\;,\\
I_{\sigma_s} &=& \int_{\Lambda} \frac{d^3k}{(2\pi)^3}\frac{1}{E_k^s}\;.\;\;
\eea
Since the steps to obtain these integrals for MSS are described in detail in previous 
references~\cite{Farias:2016let,Duarte:2018kfd,Avancini:2019ego}, here we will just show the final 
results of each of these integrals:
\begin{align}
I_{\sigma_l}^{\rm MSS} &= 2I_{\rm quad}(M_0) - (M_l^2 - M_0^2 + \Delta^2)I_{\rm log}(M_0) + I_3\nn\\
& + 3\left[\frac{\mathcal{M}^2}{4} + \mu^2(M_l^2 - M_0^2 - \mathcal{M})\right]I_1 + 2I_2,\\
I_{n_I}^{\rm MSS} &= 2\mu\Delta^2 I_{\rm log}(M_0)\nn\\
&3\mu\left[\frac{\mathcal{M}^2}{4} + \mathcal{M}(M_0^2 - M_l^2) + M_l^2\mu^2 
+ \frac{2\mu^2M_0^2}{3} \right]I_1\nn\\
& + 2\mu I_2 - \frac{5\mu M_l^2}{4}\left[3\mathcal{M}^2 + 4\mu^2M_l^2\right]I_4\nn\\
& + \frac{5\mu}{4}\left(4\mu^2(M_0^2-2M_l^2) - 3\mathcal{M}^2\right)I_5 + I_6\;,\\
I_{\sigma_s}^{\rm MSS} &= I_{\rm quad}(M_{0s}) +\frac{M_{0s}^2 - M_s^2}{2}I_{\rm log}(M_{0s}) + I_7\;,
\end{align}
where $M_{0s}$ is the vacuum strange quark mass, obtained in the $T = \mu = \Delta = 0$ limit,
and the remaining definitions,
\begin{align}
I_3 & =\frac{15}{16}\sum_{j = \pm 1}\int\frac{d^3k}{(2\pi)^3}
\int\limits_0^{\infty}\frac{t^2 dt}{\sqrt{1+t}}\nn\\
& \times \frac{1}{E_k^l} \frac{j\mu(\mathcal{M} - 2j\mu E_k^l)^3}
{\left[(k^2 + M_0^2)t + (E_k^l + j\mu)^2 + \Delta^2\right]^{\frac{7}{2}}}\;,\\
I_4 & = \int\frac{d^3k}{(2\pi)^3}\frac{1}{(k^2 + M_0^2)^{\frac{7}{2}}}\;,\\
I_5 & = \int\frac{d^3k}{(2\pi)^3}\frac{k^2}{(k^2 + M_0^2)^{\frac{7}{2}}}\;,\\
I_6 & = \frac{35}{32}\sum_{j = \pm 1}\int\frac{d^3k}{(2\pi)^3}
\int\limits_0^{\infty}\frac{t^3 dt}{\sqrt{1+t}}\nn\\
& \times \frac{jE_k^l(\mathcal{M} - 2j\mu E_k^l)^4}
{\left[(k^2 + M_0^2)t + (E_k^l + j\mu)^2 + \Delta^2\right]^{\frac{9}{2}}}\;,\\
I_7 & = \frac{3}{4}\int\frac{d^3k}{(2\pi)^3}
\int\limits_0^{\infty}\frac{t dt}{\sqrt{1+t}}\nn\\
& \times \frac{(M_{0s}^2 - M_s^2)^2}
{\left[(k^2 + M_{0s}^2)t + k^2 + M_s^2\right]^{\frac{5}{2}}}\;.
\end{align}
Note that integrals $I_1$ to $I_7$ are all finite, and must be performed up to infinity in $k$.
This is the fundamental difference between TRS, where we cut the whole integral in the 
cutoff $\Lambda$, and MSS, where all finite medium contributions are separated and performed 
for the whole momentum range.

Finally, the MSS expression for the normalized thermodynamic potential reads
\begin{align}
&\Omega_{\rm NJL}^{\rm MSS}(\sigma_l,\sigma_s,\Delta) 
= \frac{2\sigma_l^2+\sigma_s^2}{8G}+\frac{K\sigma_s\sigma_l^2}{16G^3}\nn\\
&+\left(G+\frac{K\sigma_s}{4G}\right)\frac{\Delta^2}{\left(2G+\frac{K\sigma_s}{4G}\right)^2}\nn\\
&-2N_c\Biggl\{\tilde{\mathcal{M}}I_{\rm quad}(M_0) + \frac{M_s^2 - M_{0s}^2}{4}I_{\rm quad}(M_{0s})\nn\\
& - \frac{1}{4}\left(\tilde{\mathcal{M}^2} - 4\mu^2\Delta^2\right)I_{\rm log}(M_0)
-\frac{M_s^2 - M_{0s}^2}{8}I_{\rm log}(M_{0s})\nn\\
&+\int\frac{d^3k}{(2\pi)^3}\left[\frac{\tilde{\mathcal{M}^2} - 4\mu^2\Delta^2}{4(E_{k,0}^l)^3}
- \frac{\tilde{\mathcal{M}}}{E_{k,0}^l}- 2E_{k,0}^l\right.\nn\\
& + E_k^s - E_{k,0}^s
- \frac{M_s^2 - M_{0s}^2}{2E_{k,0}^s} + \frac{(M_s^2 - M_{0s}^2)^2}{8(E_{k,0}^s)^3}\nn\\
& + \sum_{j = \pm}\sqrt{(E_k^l + j\mu)^2 + \Delta^2}\Biggl]\Biggl\}\;,
\end{align}
with the definitions $\tilde{\mathcal{M}} = \Delta^2 + M_l^2 - M_0^2$, $E_{k,0} = \sqrt{k^2 + M_0^2}$ and 
$E_{k,0}^s = \sqrt{k^2 + M_{0s}^2}$.
\section{Results and Discussions}
\label{sec3}

In this paper we have considered  the SU(3) version of 
the NJL model at finite isospin imbalance incorporating the strange quark sector and the KMT determinant for both  the zero and finite temperature cases. Thus, we have extended our previous study of the QCD equation of state at non-zero isospin density and zero temperature within the SU(2) version of the Nambu-Jona-Lasinio model. Besides, in this work we have also considered  for the SU(2) model the effects of finite temperature  in order to perform a complete
comparison between the two versions of the NJL model.

The effects of the regularization scheme are discussed in details. We have used two alternative approaches for the regularization of the 
nonrenormalizable NJL model, the first one, which we have named  TRS (Traditional Regularization Scheme), is the most frequently found in the literature where the ultraviolet divergences are regularized through a sharp 3D cutoff, as 
shown in section-\ref{sec2}. It is important to point out that the TRS approach does not disentangle finite medium contributions 
from the infinity vacuum term and physically meaningful contributions are usually discarded. The second scheme, which we named
MSS (Medium Separation Scheme), is capable of disentangling exactly the vacuum divergent term from the finite medium ones and only 
the truly divergent vacuum is regularized through a sharp 3D cutoff. It will be discussed in what follows how dependent are the observables on the chosen regularization scheme and which one is the more
appropriated for each particular situation.

The most important first principle approach to QCD in the non-perturbative regime is the lattice QCD simulation. Since for the finite isospin scenario no sign problem is found in the LQCD calculations, whenever possible, our results are compared with recent LQCD data. In order to make possible such
comparisons, we have had to fit the SU(2) and the SU(3) NJL model parameters according to the pion mass and pion decay constant adopted in LQCD calculations.
 In tables-\ref{parameter_setsSU2} and \ref{parameter_setsSU3} our 
fitted parameters are shown for the SU(2) and SU(3) NJL models  respectively. 
To compare our results with lattice simulations we used Set-I for zero temperature and Set-II for finite temperature cases.
The fitting procedure for the SU(3) version of the NJL model is more 
involved and due to this fact we have included some details of this parametrization procedure in the appendix-\ref{ap_su3}. 

Next, we discuss our results for the SU(2) and SU(3) NJL models at finite isospin density and zero and finite temperature using the 
TRS and MSS regularization schemes. At this point, we would like to emphasize the fact that the results for finite isospin and zero temperature within the SU(2) NJL model have  been obtained in our previous paper\cite{Avancini:2019ego}. However, they are shown here for a complete comparison between the NJL model versions as we have extended the zero temperature results for the SU(3) case. The finite temperature extension, however, is a completely new addition in this present work, which has been done here for both SU(2) and SU(3) cases.

\subsection{Zero Temperature results}
 
\begin{figure}
\begin{center}
\includegraphics[scale=0.33]{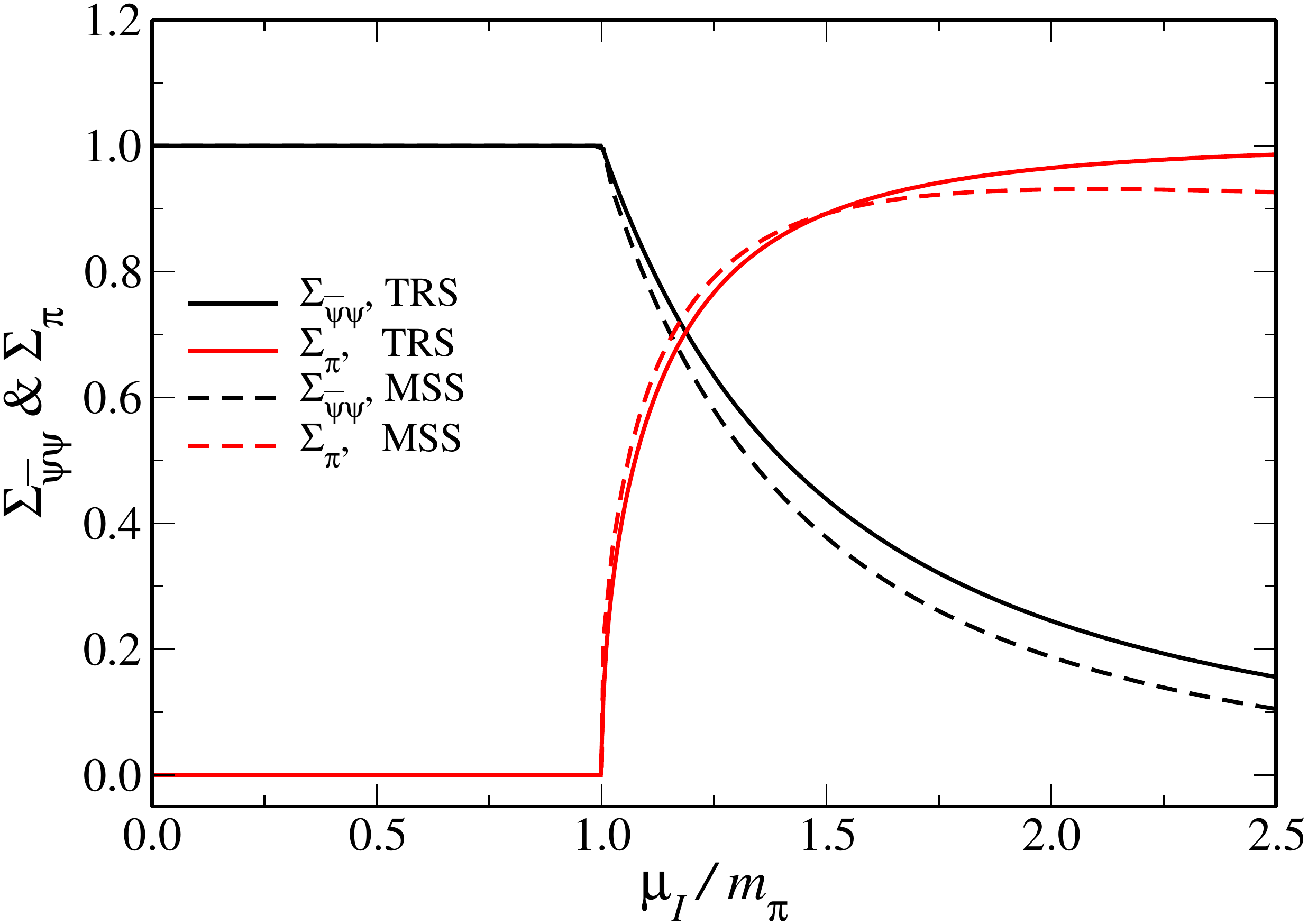}
\caption{(Color online) Chiral and pion condensates as functions of $\mu_I$ at $T = 0$, evaluated according to Eqs.~\eqref{condCh} and~\eqref{condPi}, for TRS and MSS.}
\label{figcond}
\end{center}
\end{figure}
 
We start by showing our zero temperature SU(3) results, 
Figure \ref{figcond}, for the chiral and pion condensates, which have been evaluated respectively according to the equations,
\begin{eqnarray}
 \Sigma_{\bar{\psi}\psi} & = & \frac{m_l}{m_{\pi}^2f_{\pi}^2}\left[\frac{\sigma_l - \sigma_l^0}{2G}\right] + 1, \label{condCh}\\
 \Sigma_{\pi} & = & \frac{m_l}{m_{\pi}^2f_{\pi}^2}\frac{\Delta}{2G + K\frac{\sigma_s}{4G}},\label{condPi}
\end{eqnarray}
as a function of the isospin chemical potential $\mu_I$ for both the TRS and MSS approaches. Here it is worth to mention that the chiral condensate in Eq.~\eqref{condCh} does not include the contribution from the strange quarks. We have used the same definitions for these quantities as LQCD~\cite{Brandt:2017oyy,Brandt:2018wkp}, as we have compared our results against them for the case of finite temperature\footnote{Note that there is an apparent $\frac{1}{2}$ factor difference between our definition of the condensates and the definition given in Refs~\cite{Brandt:2017oyy,Brandt:2018wkp}, like it has been done in Ref.~\cite{Adhikari:2020ufo}. This has been done to compensate for the same factor in our definitions of $\sigma_l$ and $\Delta$.}. Our results show that the difference between both approaches increases at higher $\mu_I$, where the condensates calculated in the MSS scheme are systematically lower than the corresponding ones calculated within the TRS scheme. 

\begin{figure}
\begin{center}
\includegraphics[scale=0.33]{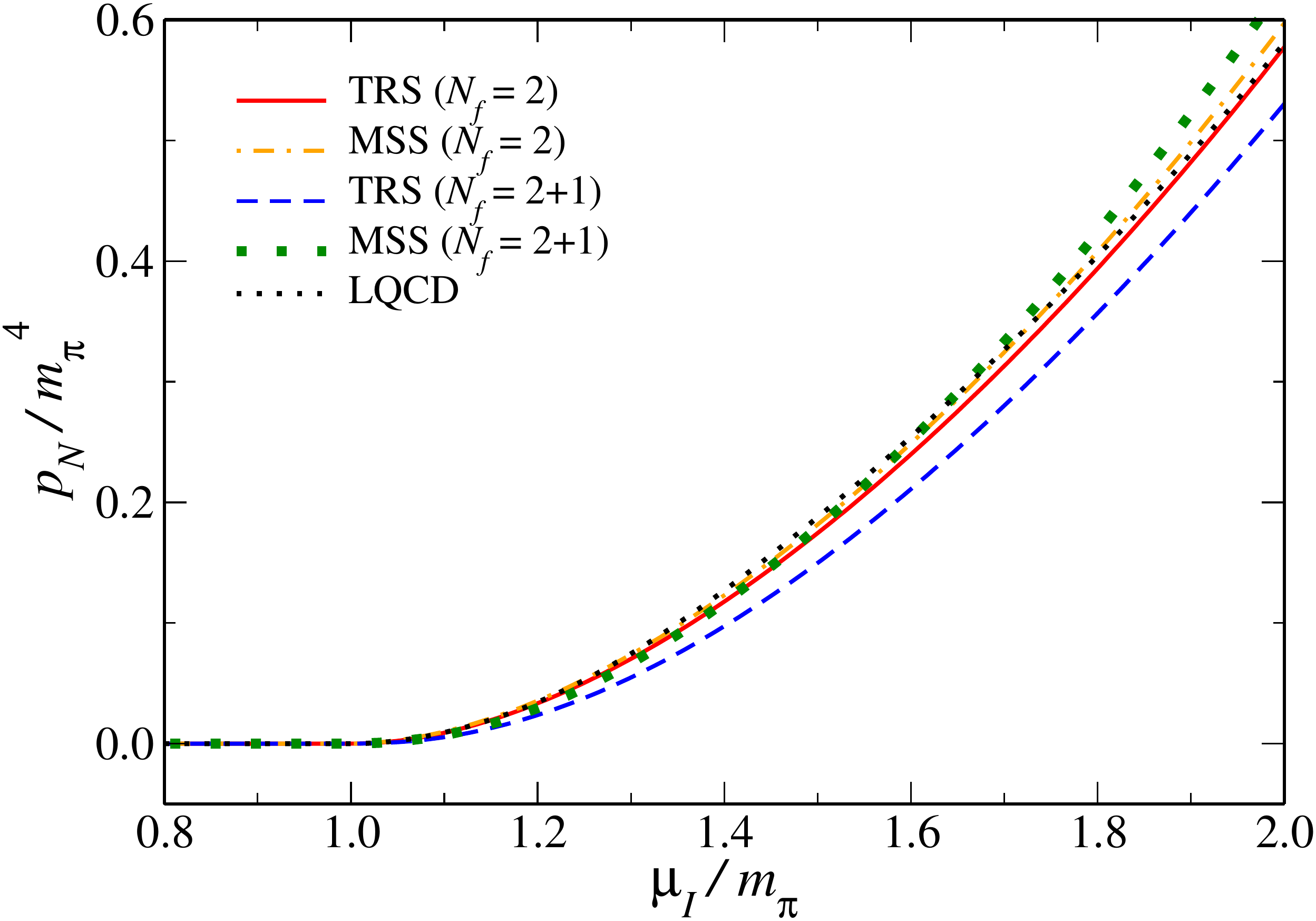}
\caption{(Color online) Normalized pressure $p_N$ as a function of $\mu_I$ at $T = 0$, for SU(2) (solid and dot-dashed lines) and SU(3) (dashed and dotted lines) comparing TRS, MSS and lattice results from Ref.~\cite{Brandt:2018bwq} (small dots).}
\label{figPzT}
\end{center}
\end{figure}

\begin{figure}
\begin{center}
\includegraphics[scale=0.33]{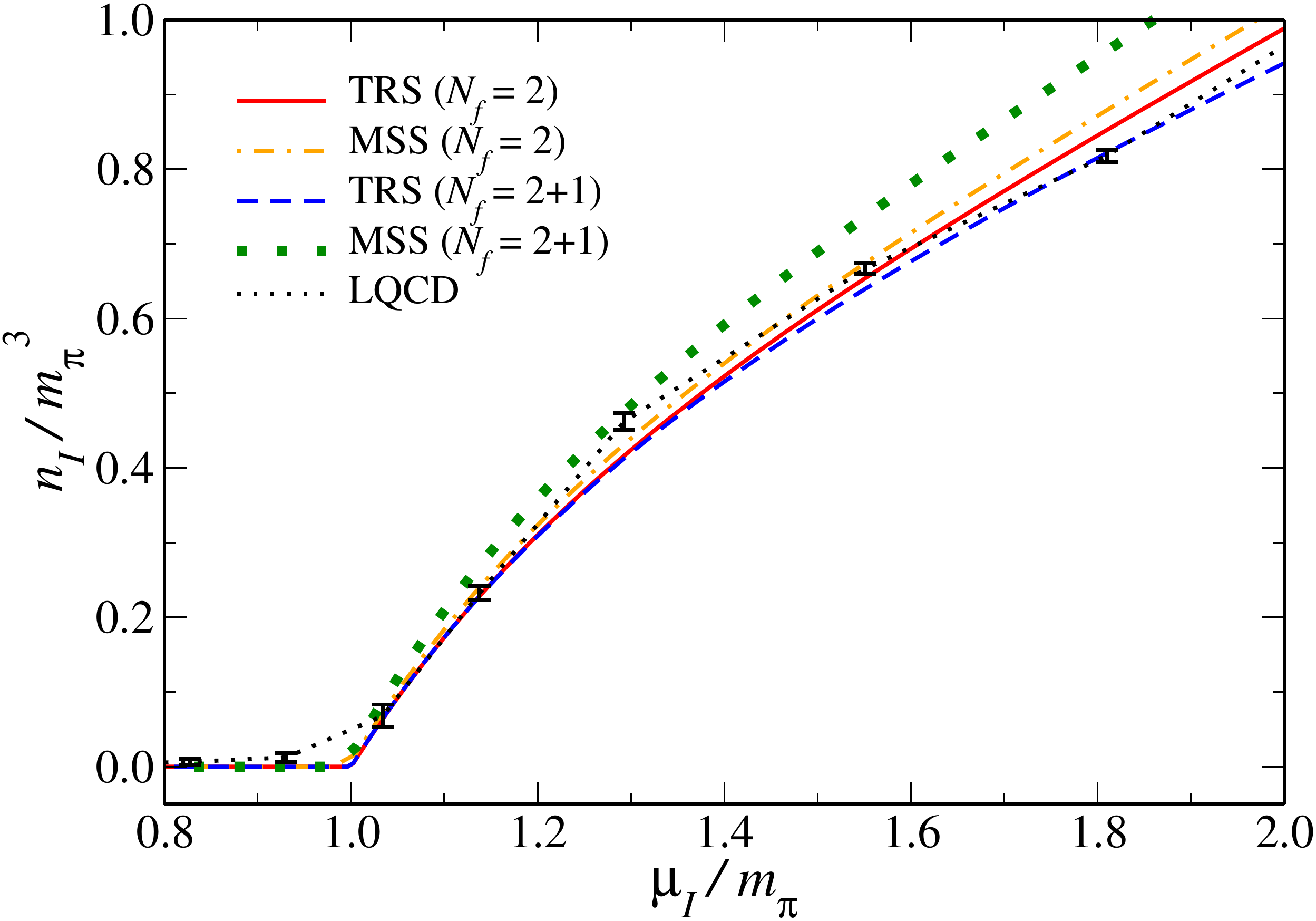}
\caption{(Color online) Isospin density $n_I$ as a function of $\mu_I$ at $T = 0$, for SU(2) (solid and dot-dashed lines) and SU(3) (dashed 
and dotted lines) comparing TRS, MSS and lattice results from Ref.~\cite{Brandt:2018bwq} (small dots).}
\label{fignIzT}
\end{center}
\end{figure}

The results for the normalized pressure are shown in Figure \ref{figPzT} as a function of the isospin chemical potential. The normalized pressure $p_N$ is defined as:
\begin{equation}
p_N = P_{\rm NJL}(T,\mu_I) - P_{\rm NJL}(T,\mu_I=0), \label{pre_normal}
\end{equation}
where $P_{\rm NJL}$ is defined in Eq.~\eqref{P_njl}. 
It is apparent of this latter figure that compared to the LQCD results~\cite{Brandt:2018bwq}, the SU(3)-NJL is better in the MSS scheme (dotted lines) than in the TRS scheme (dashed lines), at least, for $\mu_I \leq 2 m_\pi$. It is also apparent from Figure \ref{figPzT} that the SU(2)-NJL results~\cite{Avancini:2019ego} agree slightly better with LQCD data than the SU(3) ones and the differences between TRS and MSS are less important in this case. 

The results for the isospin density (see Eq.~\eqref{nI_njl}) as a function of $\mu_I$ at $T=0$ are shown in Figure \ref{fignIzT}. Although the lattice data points are few, one can see that for small $\mu_I$ the SU(3)-NJL model using the MSS scheme is in better agreement with LQCD results~\cite{Brandt:2018bwq}. Nevertheless, when $\mu_I$ increases the SU(3)-NJL model in the TRS scheme seems to be closer to the LQCD results. As compared to the SU(2)-NJL model results, the MSS scheme is closer to lattice data for $\mu_I/m_\pi < 1.6$. Thus, for both SU(2) and SU(3) versions of NJL it seems that for larger isospin chemical potential the TRS scheme is closer to lattice data. If one considers the overall trend of the lattice data, the SU(2)-NJL model seems to reproduce better the lattice data, as already noticed 
in our discussion of the behavior of the pressure. We mention here that  it would be highly desirable to get more lattice data in order to better distinguish the differences among the versions of NJL and  regularization schemes. 

\begin{figure}
\begin{center}
\includegraphics[scale=0.33]{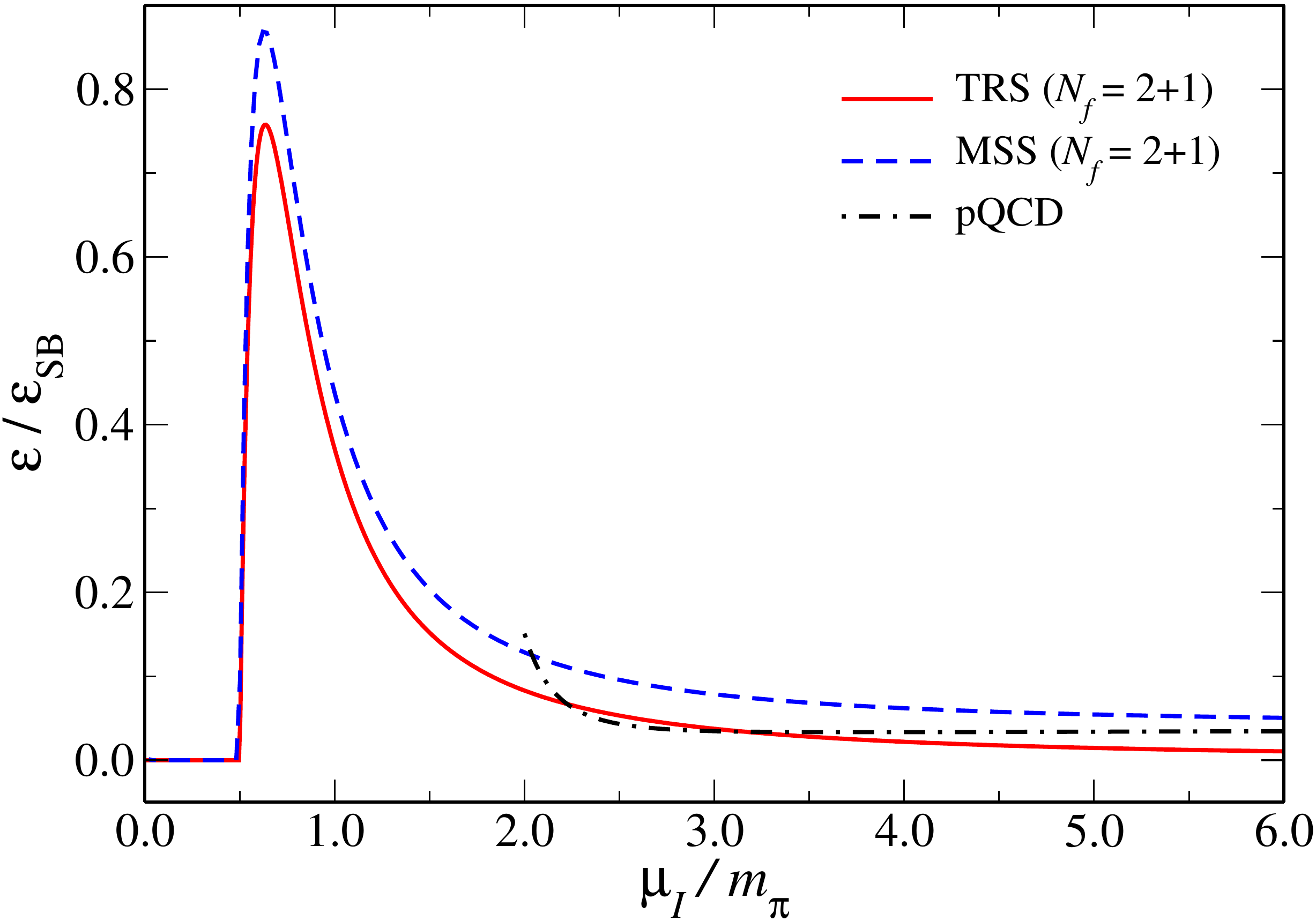}
\caption{(Color online) Comparison of the energy density scaled by the Stefan-Boltzmann limit for zero temperature and finite isospin chemical potential. The pQCD results for higher values of $\mu_I/m_\pi$ are obtained using the expressions from Ref~\cite{Graf:2015pyl}.}
\label{figedr}
\end{center}
\end{figure}

\begin{figure}
\begin{center}
\includegraphics[scale=0.33]{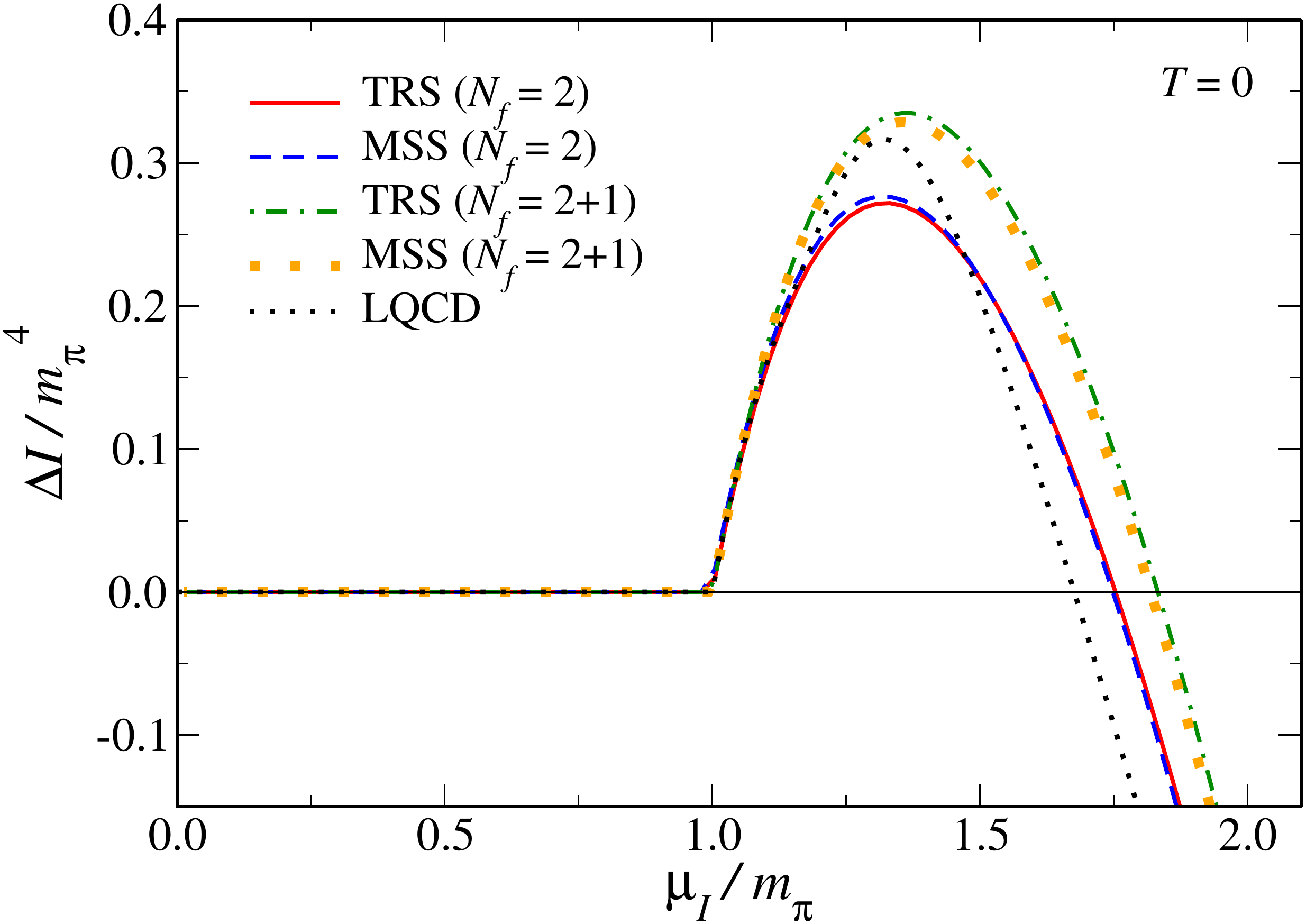}
\caption{(Color online) Normalized interaction measure ($\Delta I$) as a function of $\mu_I$ for SU(2) (solid and dot-dashed lines) and SU(3) (dashed and dotted lines) comparing TRS, MSS and lattice results from Ref.~\cite{Brandt:2018bwq,Vovchenko:2020crk} (small dots).}
\label{figtAzT}
\end{center}
\end{figure}

Next up in Figure \ref{figedr}, we show the scaled energy density (see Eq.~\eqref{nI_njl}) as a function of the scaled $\mu_I$ chemical potential for the  SU(3)-NJL model and $T=0$. The energy density is scaled with its ideal or the Stefan-Boltzmann limit for finite $\mu_I$ and $\mu_B=T=0$, given by
\begin{equation}
\varepsilon_{\text{SB}}(\mu_I) =  \frac{N_cN_f}{4\pi^2}\mu_I^4.
\end{equation}  
A comparison with the perturbative QCD calculation~\cite{Graf:2015pyl} is performed for larger values of $\mu_I$\footnote{We would like to mention that in this reference~\cite{Graf:2015pyl}, $\mu_I$ is defined as $\mu_I =\mu_u-\mu_d$, whereas in our case $\mu_I = \frac{\mu_u-\mu_d}{2}$. Hence, we have rescaled our expression accordingly, by a factor 2.}. Some care is necessary when using NJL for larger $\mu_I$, since we have the $\Lambda$-cutoff as a natural 
momentum scale which makes the model quantitatively trusty for 
$\mu_I m_\pi \leq 3 \sim 4$, nevertheless, only qualitative conclusions can be done extrapolating such limit. However, it is 
clear from the latter figure that the MSS scheme follows the trend of pQCD for higher values of $\mu_I$. Moreover, one clearly sees in Figure \ref{figedr} that the Stefan-Boltzmann limit is expected to be achieved only in the MSS scheme.

We finish our discussions on the zero temperature results showing in Figure \ref{figtAzT} the normalized interaction measure $\Delta I$ as a function of the scaled isospin chemical potential. The normalized interaction measure $\Delta I$ is defined as:
\begin{equation}
\Delta I = I_{\rm NJL}(T,\mu_I) - I_{\rm NJL}(T,\mu_I=0), \label{tA_normal}
\end{equation}
where $I_{\rm NJL}$ is defined in Eq.~\eqref{tA_njl}. As can be seen from Figure \ref{figtAzT}, the difference between the TRS (solid and dashed lines) and MSS (dot-dashed and dotted lines) results for both SU(2) and SU(3) are small, but the MSS results appear closer to the LQCD results~\cite{Brandt:2018bwq,Vovchenko:2020crk}. 

At zero temperature, we observe the general trend that the agreement between lattice and NJL results is better for two-flavor at larger values of $\mu_I$ and for three-flavor at lower values of $\mu_I$. Similar observations can also be found in recent $\chi$PT studies~\cite{Adhikari:2019mlf,Adhikari:2020ufo}. In the next subsection we discuss the results at finite temperature and isospin chemical potential. 

\subsection{Finite temperature results}

\begin{figure}
\begin{center}
\includegraphics[scale=0.33]{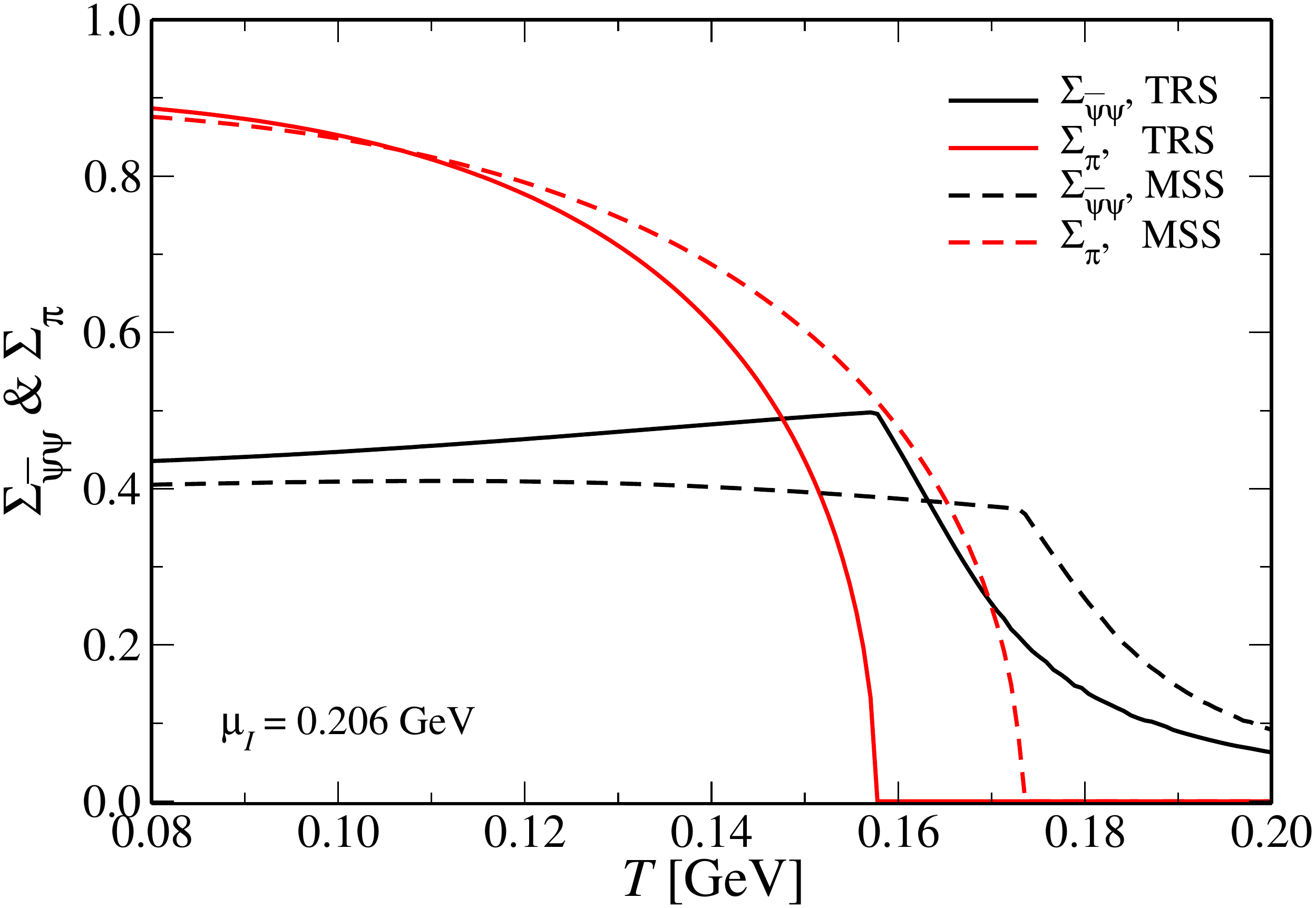}
\caption{(Color online) SU(3) Chiral and pion condensates as functions of the temperature $T$ with $\mu_I = 0.206$ GeV for TRS and MSS.}
\label{figcondsu3}
\end{center}
\end{figure}

\begin{figure}
\begin{center}
\includegraphics[scale=0.33]{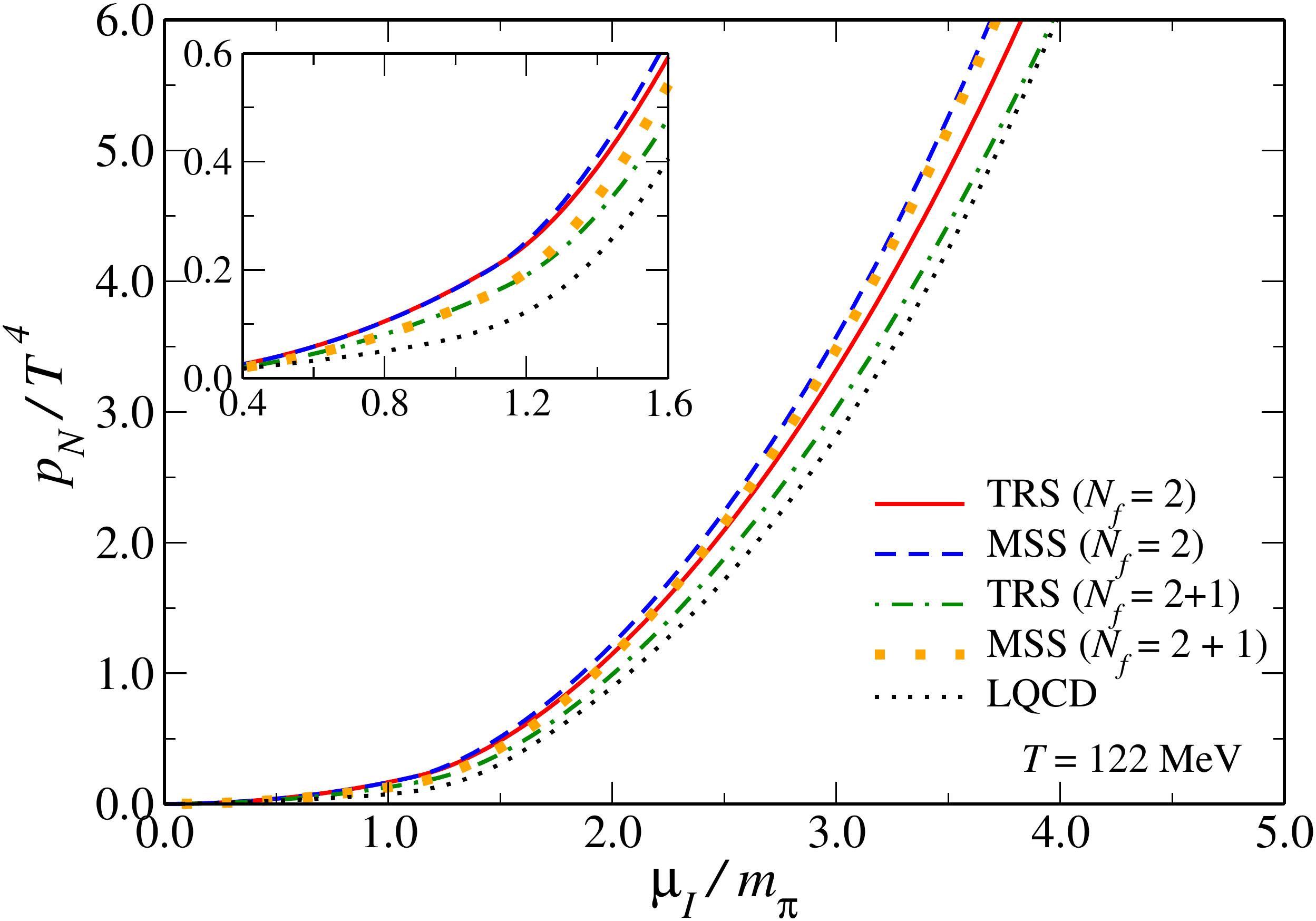}
\includegraphics[scale=0.33]{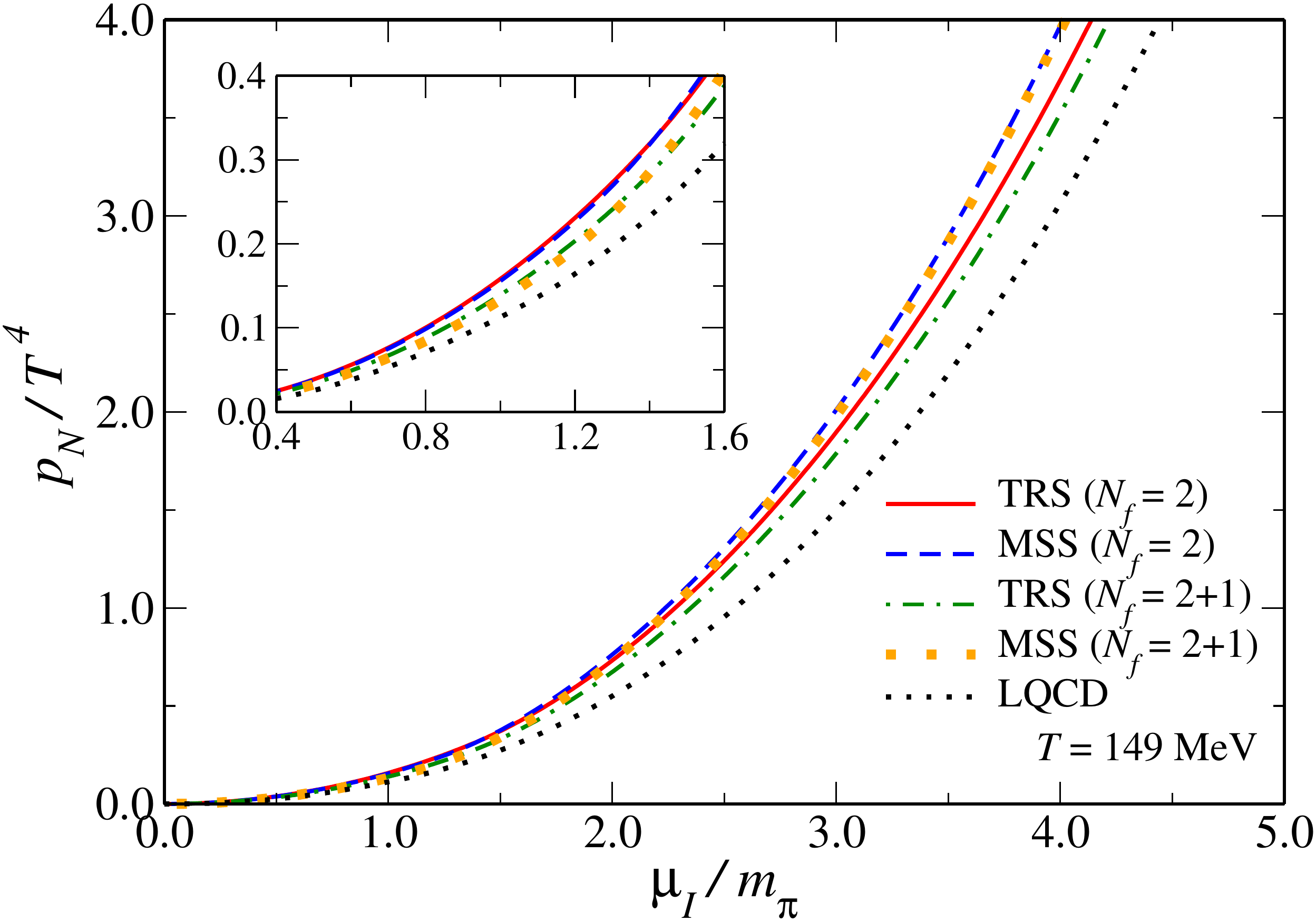}
\includegraphics[scale=0.33]{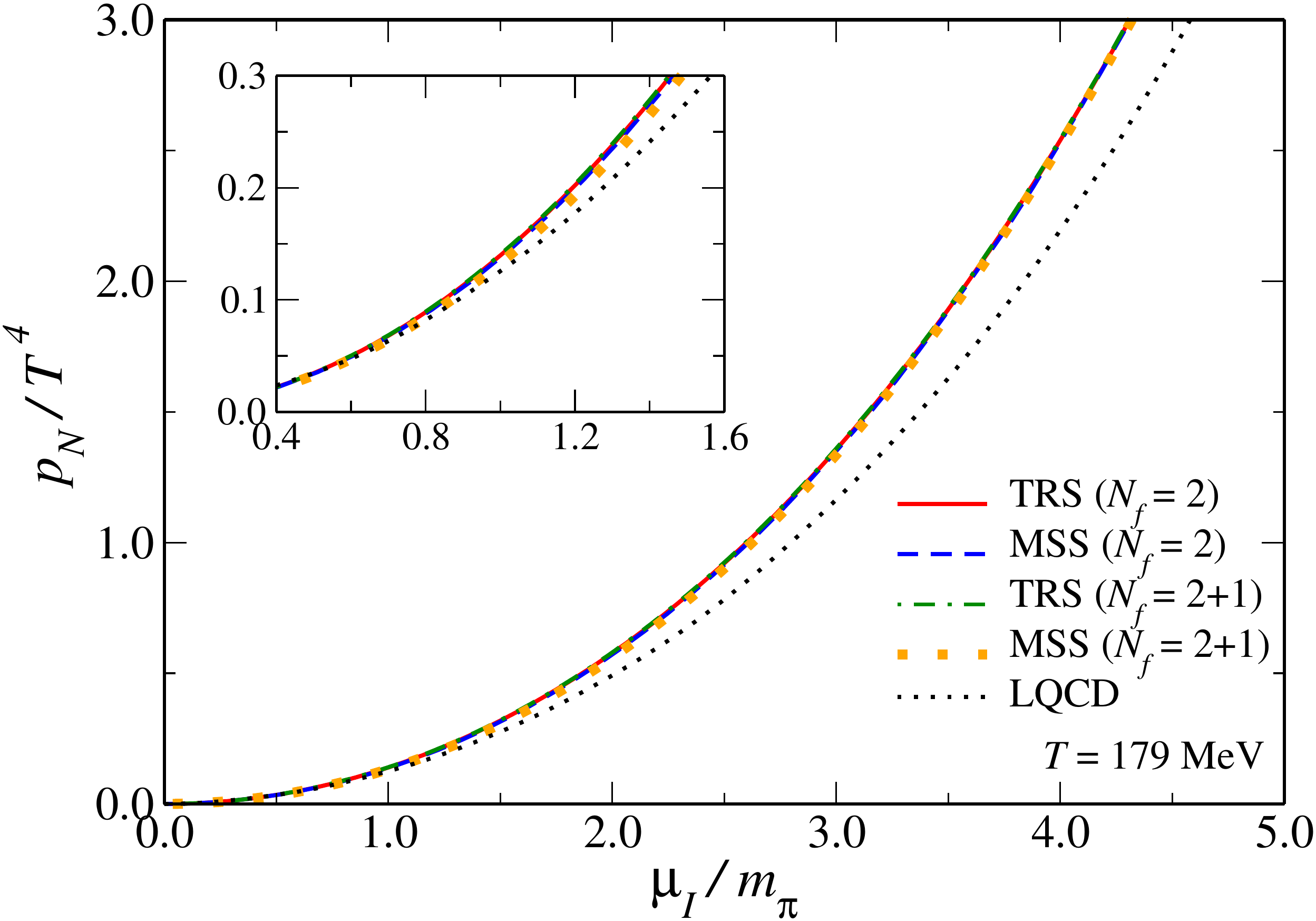}
\caption{(Color online) Normalized pressure $p_N$ as a function of $\mu_I$ for different temperatures $T$, for TRS and MSS, comparing SU(2) and SU(3) to lattice results from Refs.~\cite{Brandt:2017oyy,Brandt:2018wkp,Brandt:2019idk,Brandt:pc}.}
\label{figPfT}
\end{center}
\end{figure}

\begin{figure}
\begin{center}
\includegraphics[scale=0.33]{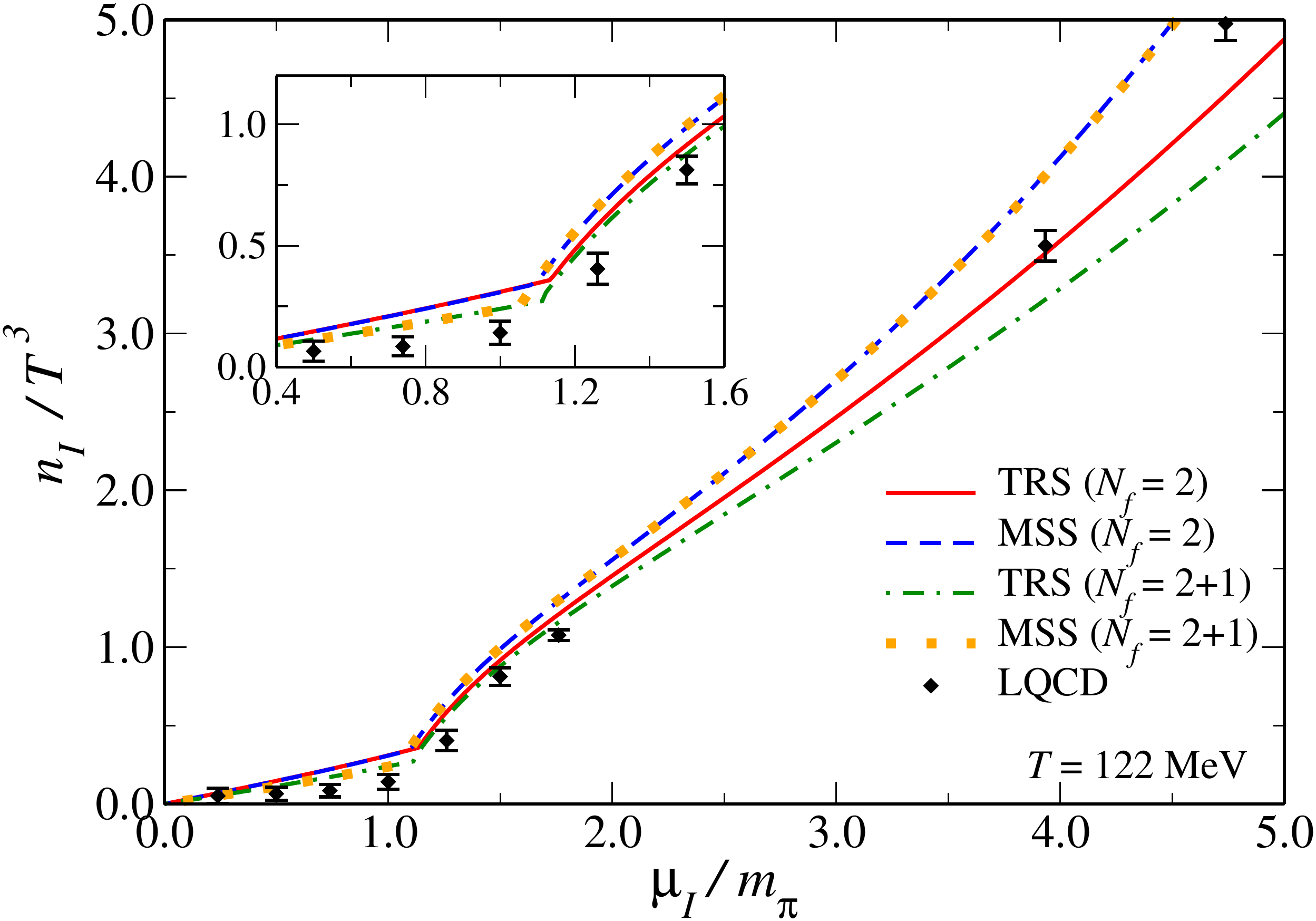}
\includegraphics[scale=0.33]{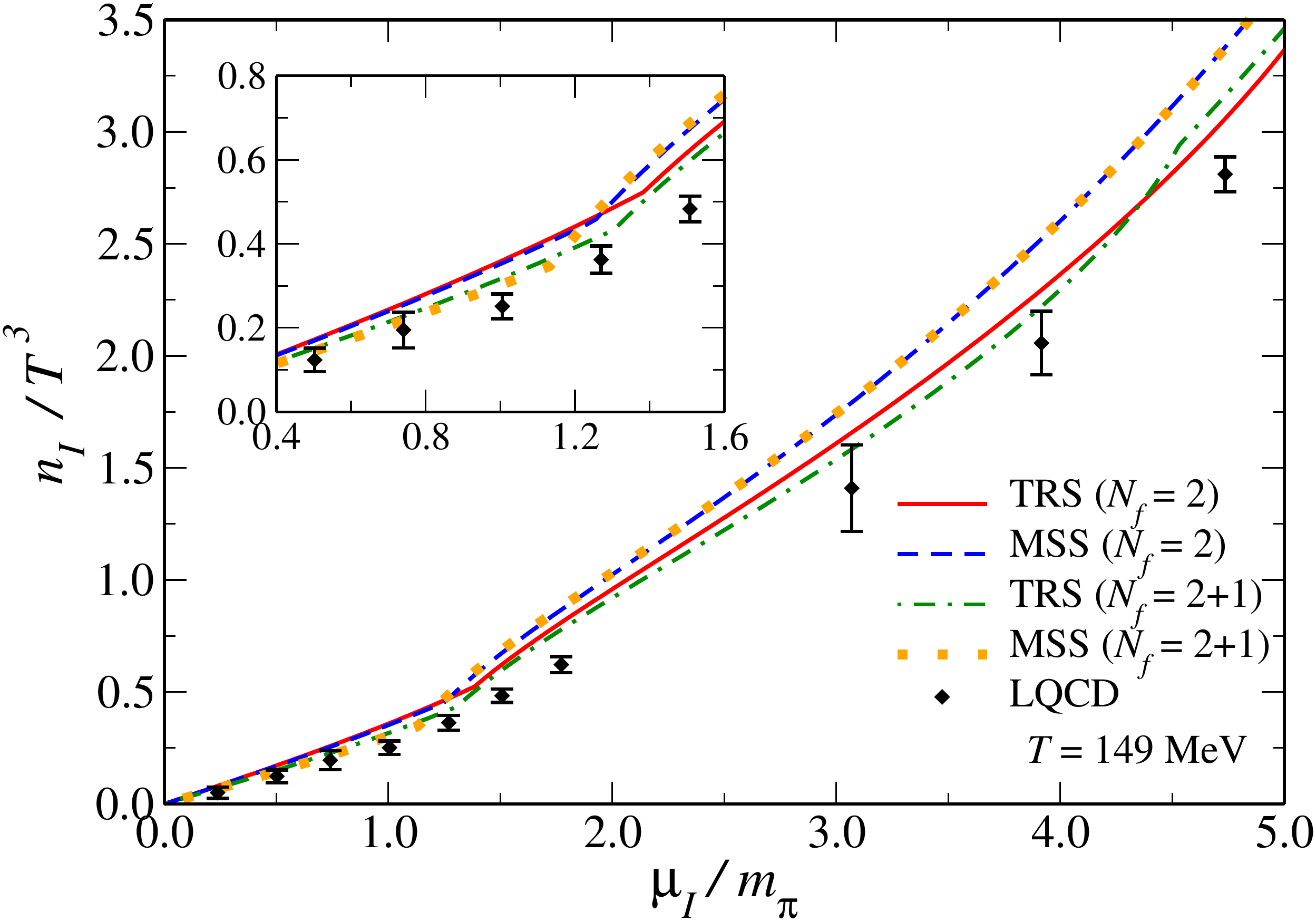}
\includegraphics[scale=0.33]{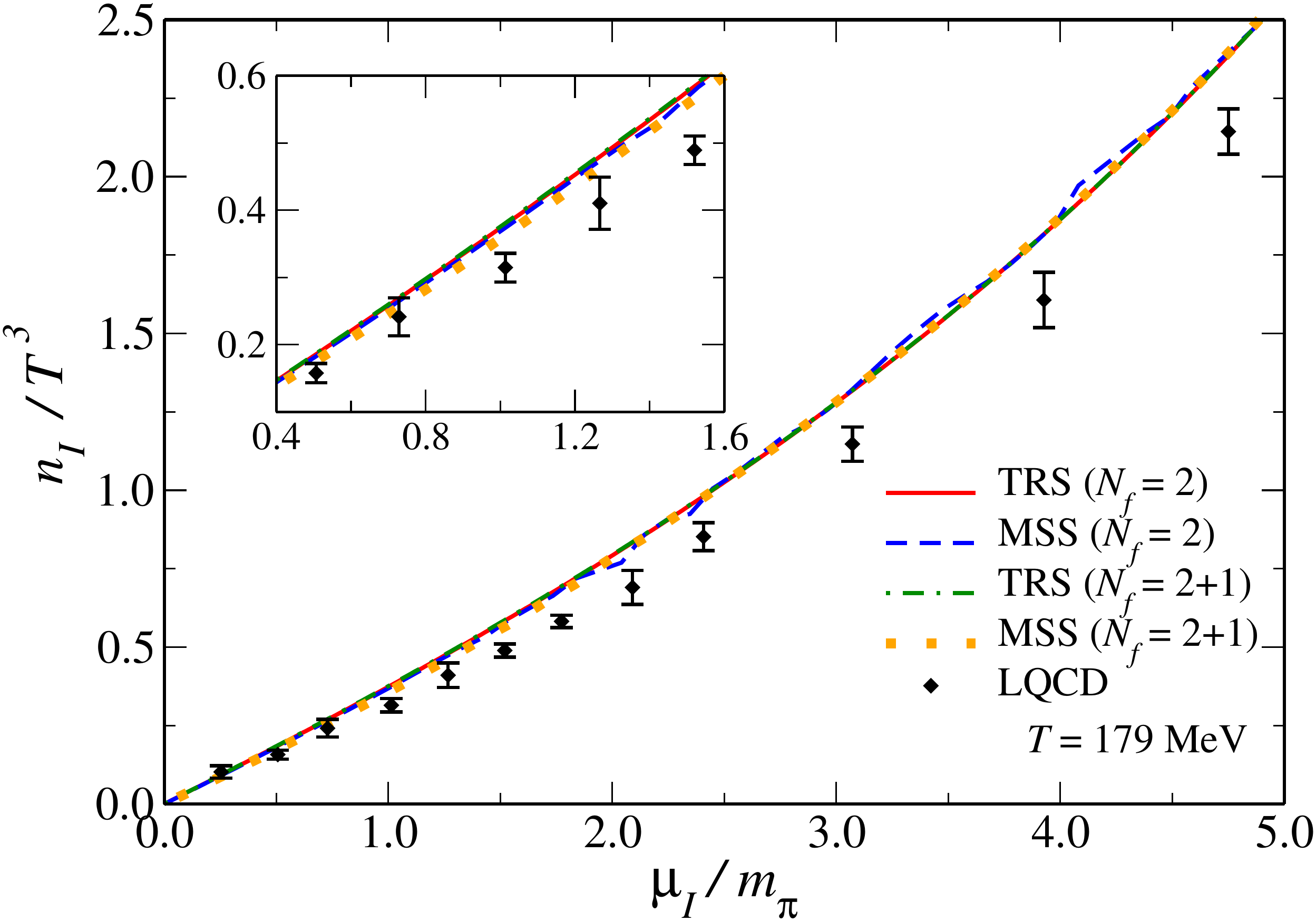}
\caption{(Color online) Isospin density $n_I$ as a function of $\mu_I$ for different temperatures $T$, for TRS and MSS, comparing SU(2) and SU(3) to lattice results from Refs.~\cite{Brandt:2017oyy,Brandt:2018wkp,Brandt:2019idk,Brandt:pc}.}
\label{fignIfT}
\end{center}
\end{figure}

\begin{figure}
\begin{center}
\includegraphics[scale=0.33]{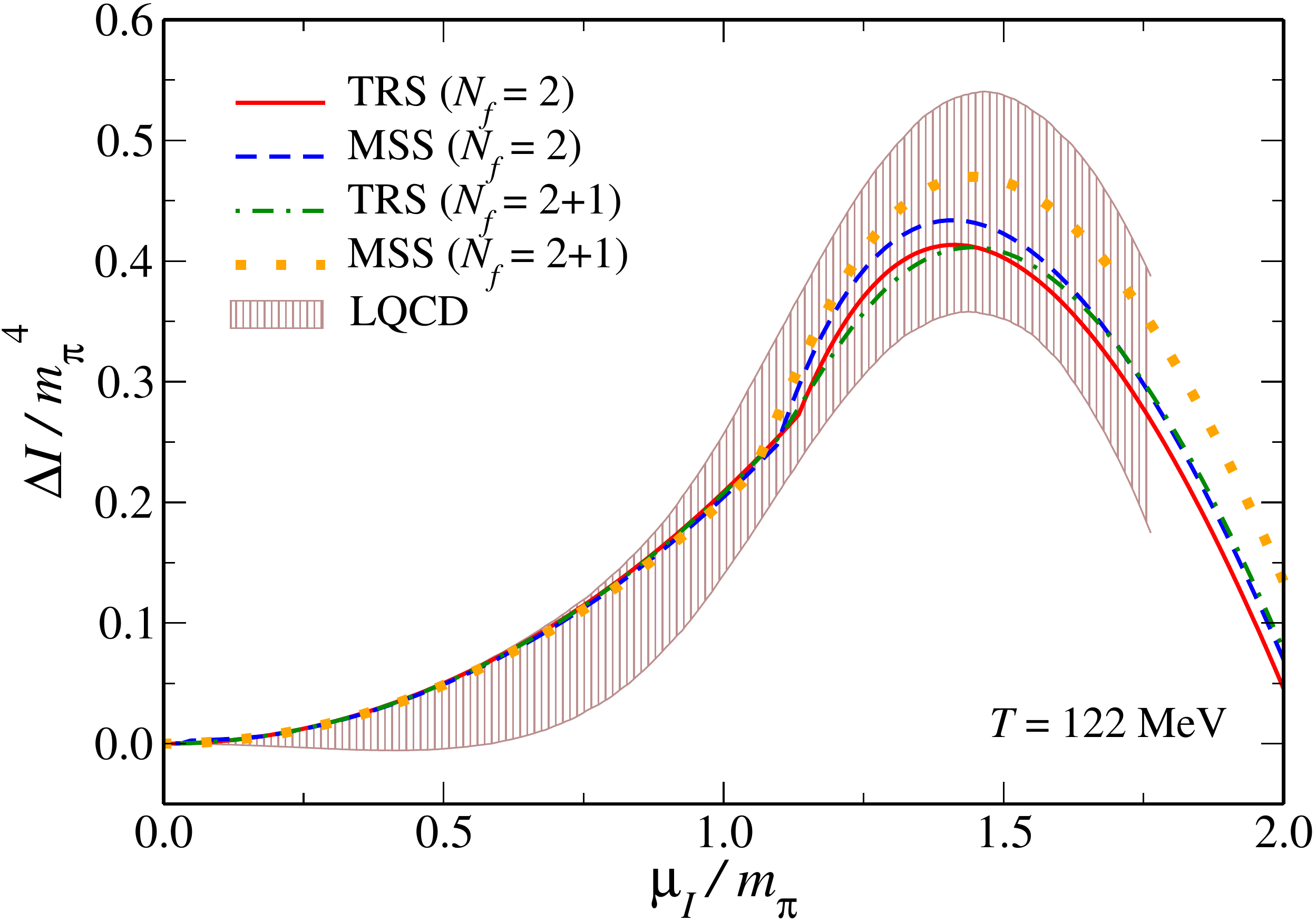}
\includegraphics[scale=0.33]{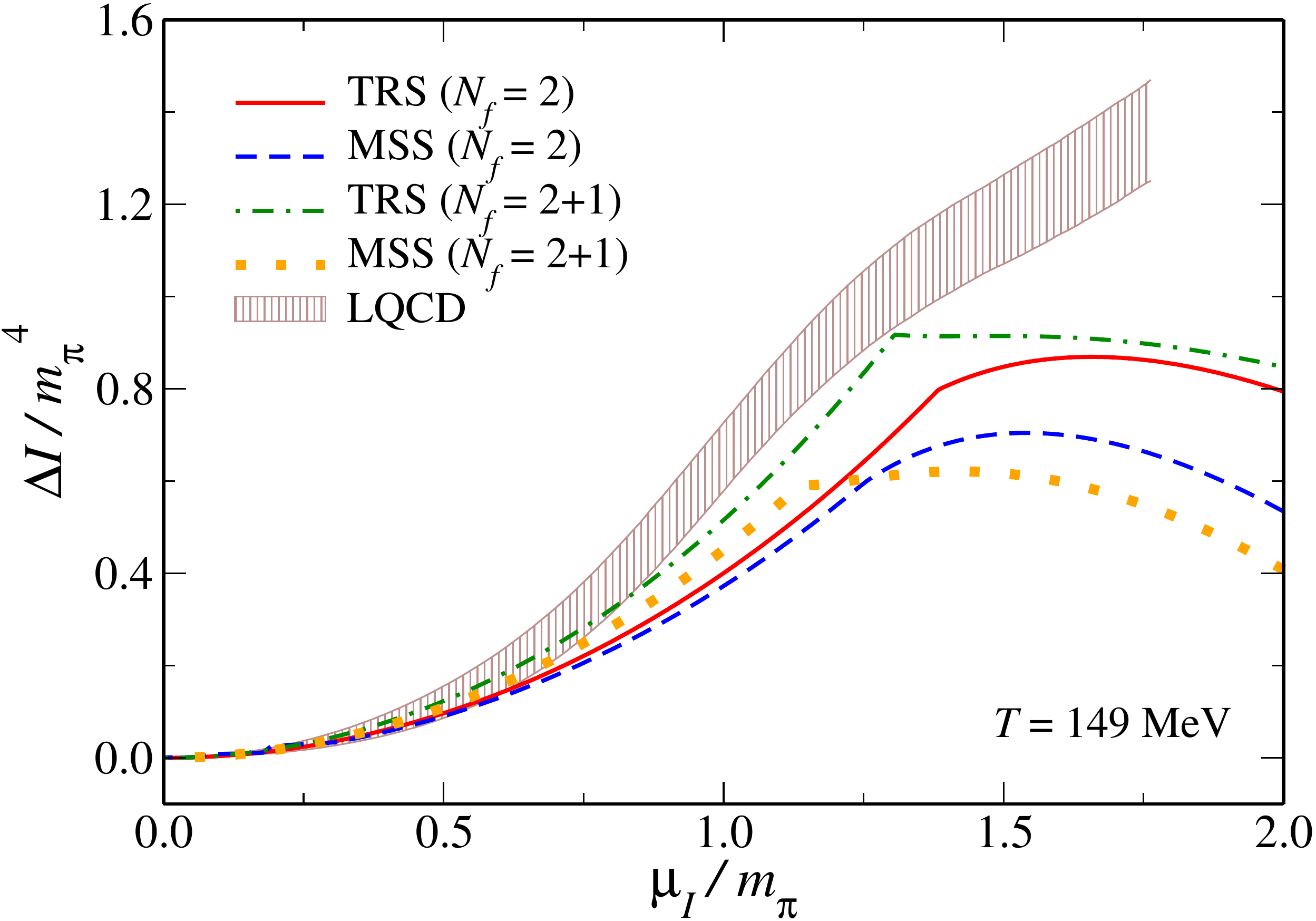}
\includegraphics[scale=0.33]{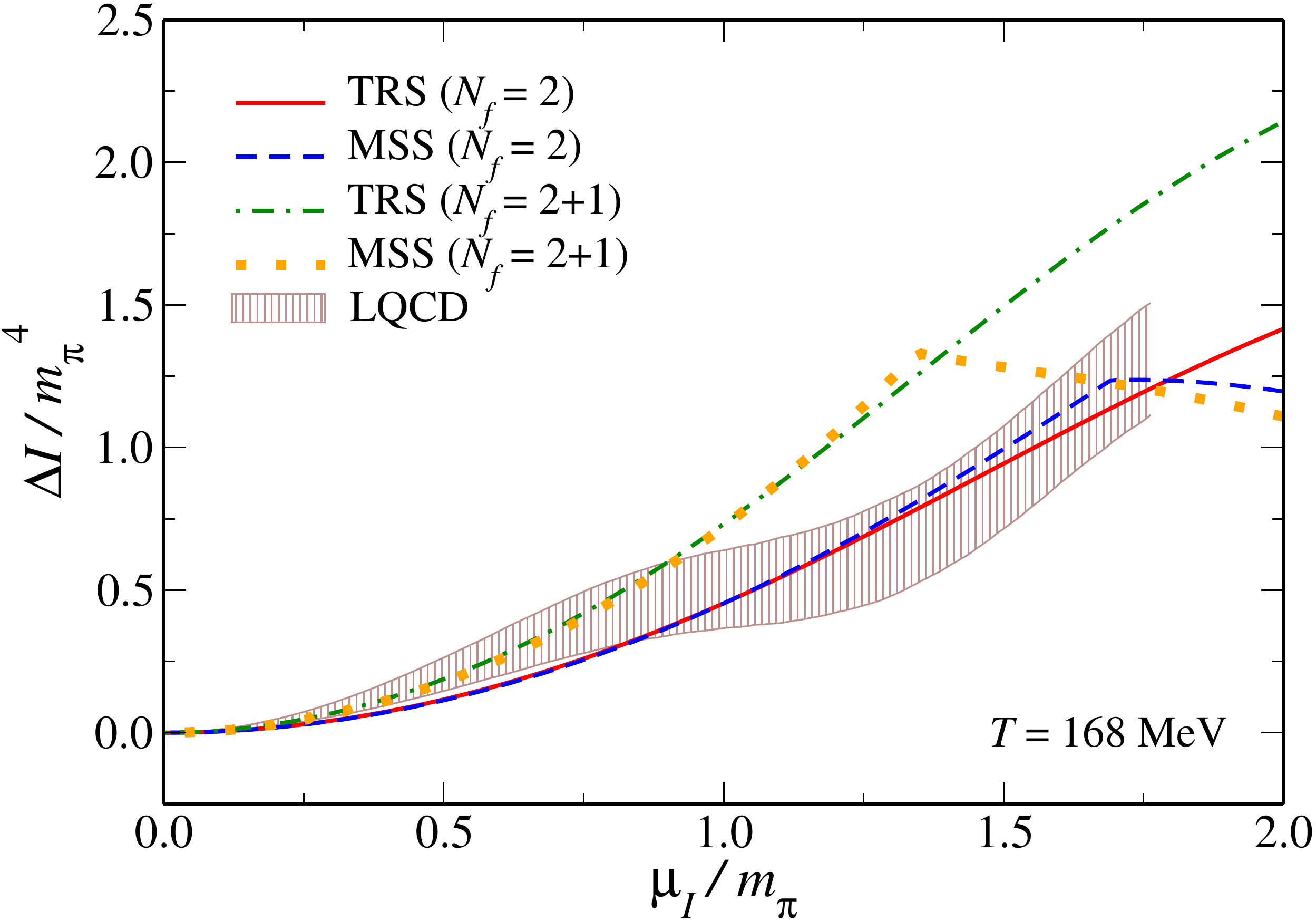}
\caption{(Color online) Normalized trace anomaly or interaction measure ($\Delta I$) as a function of $\mu_I$ for different temperatures $T$, for TRS and MSS, comparing SU(2) and SU(3) to lattice results from Ref.~\cite{Vovchenko:2020crk}.}
\label{figtAfT}
\end{center}
\end{figure}

At finite temperature we begin with Figure \ref{figcondsu3}, where the chiral and pion condensates as a function of the temperature calculated at fixed isospin chemical potential ($\mu_I=0.206$ GeV) are shown. In the latter figure   
results for the SU(3)-NJL model using both the MSS and TRS schemes are compared. One clearly sees that the melting temperature is larger in the MSS 
scheme compared to the TRS one for both condensates. However, the pion condensate melting is a second order phase transition with the critical temperature $\sim$158 MeV for TRS and $\sim$173 MeV for MSS. On the other hand, the corresponding chiral condensate behavior signals a crossover. 
Below a certain temperature (around $\sim 165$ MeV for $\mu_I=206$ MeV) the TRS result for the chiral condensate supersedes the MSS result, whereas after that temperature the TRS result decreases more rapidly than the MSS result. 

Next we discuss about our findings for the thermodynamic quantities. In figure \ref{figPfT} the results are shown for the normalized pressure, 
Eq.(\ref{pre_normal}), as a function of the  isospin chemical potential for three different temperatures, which have been chosen because they are available in LQCD simulations~\cite{Brandt:2017oyy,Brandt:2018wkp,Brandt:2019idk,Brandt:pc}. The results have been shown for both SU(2) and SU(3)-NJL model within the TRS and MSS schemes. For larger values of $\mu_I$, the SU(3)-NJL model within the TRS scheme is closer to the LQCD results, although the agreement is better for lower temperatures. 
For each of the plots we have also displayed an inset, where we show the results for smaller values of $\mu_I$ ($0.4 m_\pi <\mu_I < 1.6 m_\pi$) and in this case we observe that the MSS and TRS schemes are closer to each other, and for the largest temperature considered in this work the SU(3)-NJL using the MSS scheme is slightly better. Here, we notice a different behavior for finite temperature when compared to the zero temperature case.  For $T=0$ the SU(2)-NJL results for the $p_N$ are in general in better agreement with LQCD results (for TRS and MSS) than the SU(3) NJL model.

In Figure~\ref{fignIfT}, the isospin density as a function of isospin chemical potential is plotted for the same three temperatures as discussed above. We can observe a change of slope in the isospin density for $T = 122$ MeV and $T = 149$ MeV which is related to the formation of pion condensate, when $\Delta$ become nonzero. This change is not present in the curves correspondent to $T = 179$ MeV because this temperature is high enough to prevent the pion condensate to be formed. In this case we have a similar behavior as obtained for the normalized pressure, i. e., the SU(3)-NJL with the TRS scheme gives a better overall agreement with the LQCD results~\cite{Brandt:2017oyy,Brandt:2018wkp,Brandt:2019idk,Brandt:pc}. For the lower temperature ($T=122$ MeV) considered here, the SU(2)-NJL in TRS scheme gives good results. As before, the inset plots show that the MSS and TRS are very similar for low isospin chemical potential and the SU(3)-NJL is the better model.  At this point we once again stress the fact that the LQCD data for the isospin density is not conclusive enough because of its scarcity and oscillating nature.

Within the thermodynamic quantities we have also shown the variation of the normalized interaction measure $\Delta I$ (see Eq. \eqref{tA_normal}) as a function of isospin chemical potential for three different temperatures in Figure \ref{figtAfT}. In this case though, instead of $T=179$ MeV, the highest temperature we choose is $T=168$ MeV, based on the availability of LQCD results~\cite{Vovchenko:2020crk}. For lower temperature ($T=122$ MeV) it is apparent from Figure \ref{figtAfT} that both TRS and MSS results for SU(2) and SU(3) fall within the domain prescribed by LQCD. At $T=149$ MeV, for very low $\mu_I$, all the NJL results fall within the LQCD domain, however with increasing $\mu_I$, only SU(3) TRS result appears closer to LQCD result. For the higher temperature considered in Figure \ref{figtAfT} ($T=168$ MeV), we can see that the SU(2)-NJL results for TRS and MSS are more in agreement with the LQCD results. It is worth to mention here that for $T = 168$ MeV at higher values of $\mu_I$ the pion condensate is zero in TRS, which is not the case of MSS. The presence of a nonzero $\Delta$ makes both pressure and energy density smaller, and the change of slope is related to the isospin density, as we discussed before (see Figure~\ref{fignIfT}).

\begin{figure*}
\begin{center}
\includegraphics[scale=0.33]{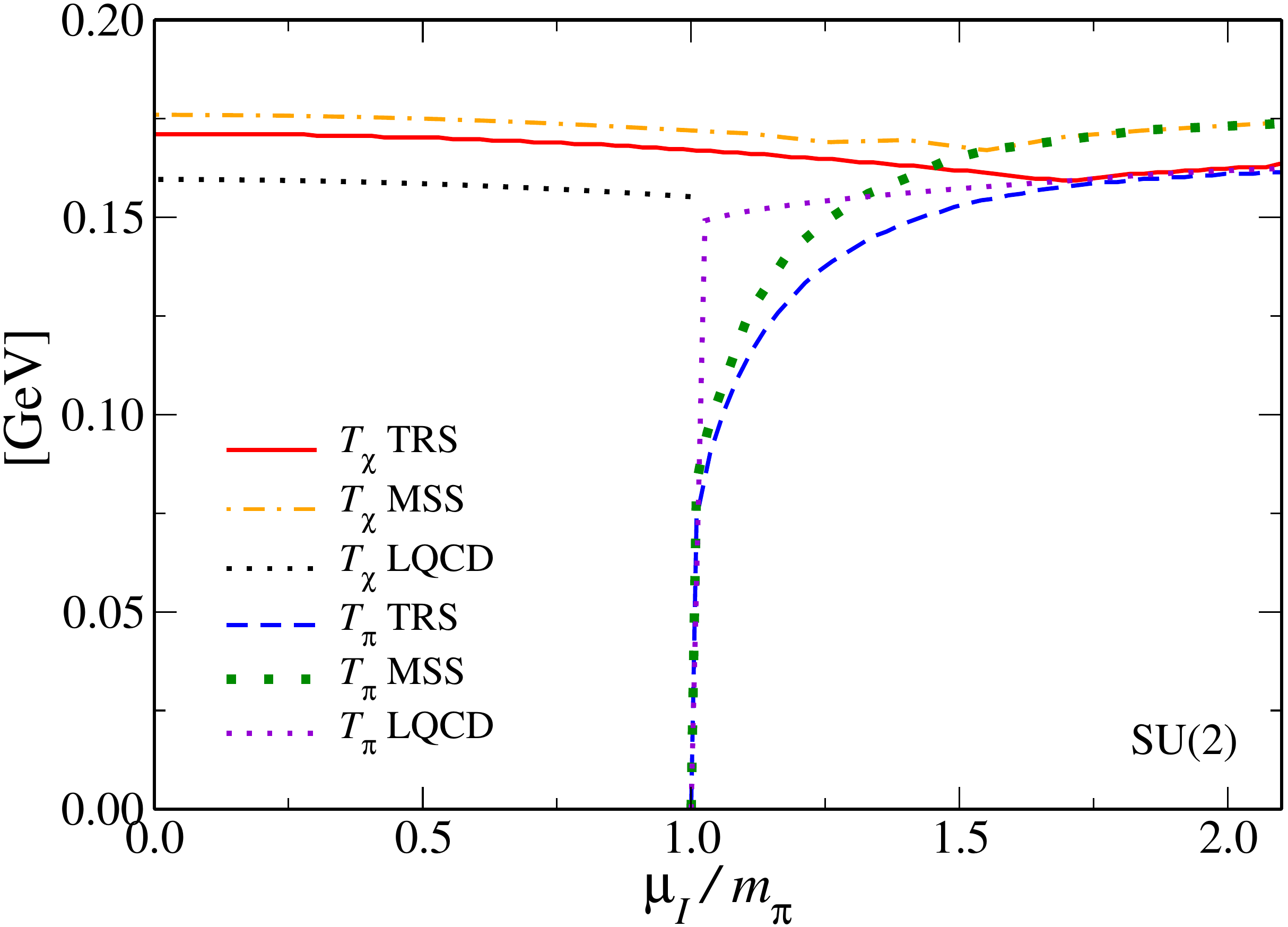}
\includegraphics[scale=0.33]{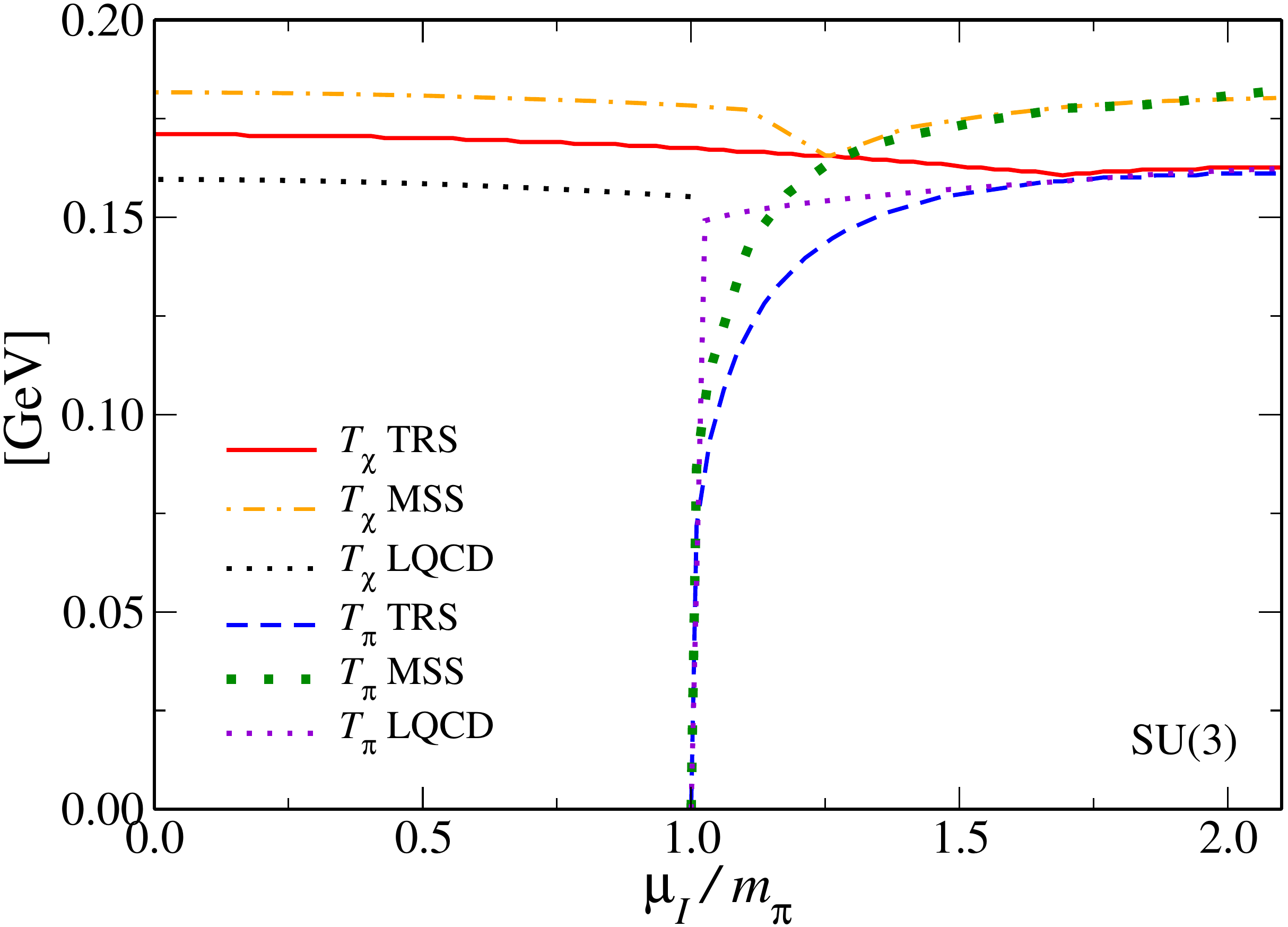}
\caption{(Color online) $T-\mu_I$ phase diagram within NJL model, implementing both TRS and MSS and comparing SU(2) and SU(3) to lattice results from Refs.~\cite{Brandt:2017oyy,Brandt:2018wkp,Brandt:2019idk,Brandt:pc}.}
\label{figphdiag}
\end{center}
\end{figure*}

We conclude our discussion for finite temperature with the $T-\mu_I$ phase diagrams shown in Figure \ref{figphdiag} for both SU(2) and SU(3). In each of the phase diagrams we have considered TRS and MSS results within NJL and compared them with the available LQCD results~\cite{Brandt:2017oyy,Brandt:2018wkp,Brandt:2019idk,Brandt:pc}. We notice that for both SU(2) and SU(3), TRS results appear closer to LQCD for pion condensation and chiral crossover, compared to MSS. For $\mu_I = 0$, the critical temperature for the chiral phase transition within NJL is higher than the same within LQCD, with maximum quantitative difference of $\sim 16$ MeV for SU(2) and $\sim 22$ MeV for SU(3). On the other hand, investigating the pseudo triple point, beyond which both the phase transitions coincide, we find that the pseudo triple points within MSS are closer to the LQCD results compared to TRS, specially for SU(3)-NJL.

%
%
%
%
%
%
%
%

\section{Acknowledgements}
We thank Gergely Endrodi and Bastian B. Brandt for useful discussions and also for providing the necessary lattice datasets, both at zero and finite temperatures. This work was partially supported by Conselho Nacional de Desenvolvimento Cient\'{i}fico e Tecnol\'{o}gico (CNPq)
under Grants No. 304758/2017-5 (R.L.S.F), No. 304518/2019-0 (S.S.A) and No. 136071/2018-0 (B. S. L.); 
as a part of the project INCT-FNA (Instituto Nacional de Ci\^encia e Tecnologia - F\'{\i}sica Nuclear
e Aplica\c c\~oes) No. 464898/2014-5 (S.S.A); U.S. DOE under Grant No. DE-FG02-00ER41132 and 
Funda\c{c}\~ao de Amparo \`a Pesquisa do Estado de S\~ao Paulo (FAPESP)
under Grant No. 2017/26111-4 (D.C.D); and Funda\c{c}\~ao de Amparo \`a Pesquisa do Estado do Rio Grande do Sul (FAPERGS), Grants No. 19/2551- 0000690-0 and No. 19/2551-0001948-3 (R.L.S.F.). D.C.D. acknowledges the support of the Simons Foundation under the Multifarious Minds Program Grant No. 557037. A.B. acknowledges the support from Guangdong Major Project of Basic and Applied Basic Research No. 2020B0301030008 and Science and Technology Program of Guangzhou Project No. 2019050001.

\appendix 

\section{NJL SU(3) Parameterization}
\label{ap_su3}

In this appendix the procedure used for the parametrization of the NJL SU(3) model is discussed. Although 
we use standard techniques  the complete expressions are not usually 
given in a complete and friendly way in the literature\cite{reviews}. Moreover, having in mind to allow
the reader to reproduce our calculations, the essential expressions are given here. In order to restrict 
the number of free parameters,  we set the u and d current quark masses equal, i.e., $m_u=m_d$. Therefore, 
we obtain by solving the self-consistent gap equations that 
the corresponding constituent u and d  quark masses are also equal to each ohter: $M_u=M_d$.  The free model parameters
are the coupling constants $G$ and $K$, the current s quark mass
$m_s$ and the cutoff parameter $\Lambda$. The value of the $m_u$=$m_d$ current quark mass is arbitrarily 
fixed. The observables that are used in the fitting procedure are the pion mass, the pion decay constant,
the kaon mass and the $\eta^\prime$ meson mass.

\subsection{Gap equations}
 The gap equations for the NJL SU(3) model are given by 
 \bea
 M_i &=& m_i - 4G \sigma_i + 2K \sigma_j\sigma_k, 
 \eea
with (ijk) being any cyclic permutation of (u,d,s), and the condensate is defined as
\bea
\sigma_i = \langle \bar{q}_i q_i \rangle = -i {\rm Tr} \left[ S_i (p) \right] = 
    -i {\rm Tr} \frac{1}{\slashed{p}-M_i}   ~ . \nonumber 
\eea
After the explicit calculation of the trace in Dirac and color spaces one obtains:
\bea
\sigma_i = \langle \bar{q}_i q_i \rangle = -4M_i I_1^i,
\eea
with 
\begin{equation}
I_1^i = \frac{N_c}{4\pi^2}\int\limits_0^\Lambda \frac{p^2dp}{E_i}
= \frac{N_c}{8\pi^2}\left[\Lambda\epsilon^\Lambda_i - M_i^2 
\ln\left(\frac{\Lambda+\epsilon^\Lambda_i}{M_i}\right) \right] ~,
\label{eqI1}
\end{equation}
where $E_i = \sqrt{p^2+M_i^2}$ and $\epsilon^\Lambda_i=\sqrt{\Lambda^2+M_i^2}$
, i=(u,d,s).

\subsection{Pion and Kaon masses and decay constants}

The dispersion relations for the pion and kaon masses are given by 
\bea
 1-2G_\pi\Pi^p_{\pi}(Q^2)\Big\vert_{Q^2=m_\pi^2} &=& 0,   \\
 1-2G_K\Pi^p_{K}(Q^2)\Big\vert_{Q^2=m_K^2} &=& 0,
 \eea
where $\Pi^p_{\pi} =\Pi^p_{uu}+\Pi^p_{dd}$=$2\Pi^p_{uu}$ and  $\Pi^p_{K}=2\Pi^p_{us}$ are
the pseudo-scalar polarization loops
for the pion and kaon mesons respectively, which can be evaluated from the general 
expressions  
\bea
\Pi_{ij}^p(Q^2) = 2\Big((I_1^i+I_1^j)-\left[Q^2-(M_i-M_j)^2\right]I_2^{ij}\Big) ~,
\label{Pi_ij}
\eea
with $I_1^i$ already given in Eq.~\eqref{eqI1} and $I_2^{ij}$ is given by
\bea
I_2^{ij}(Q^2) = \frac{N_c}{4\pi^2}\int\limits_0^\Lambda \frac{p^2 ~dp}{E_iE_j}~~ \frac{E_i+E_j}{Q^2-(E_i+E_j)^2},
\label{eqI2}
\eea
 whereas the modified couplings $G_\pi$ and $G_K$ for pions and kaons respectively are given by,
\bea
G_\pi &=& G - \frac{1}{2} K\sigma_s, \\
G_K &=& G - \frac{1}{2}K\sigma_u.
\eea 
For the pion decay constant we have
\bea
f_\pi = g_{\pi\bar{q}q} \frac{Q_\mu}{Q^2} i N_c \int d^4 p ~{\rm Tr} 
\left[ \gamma^\mu \gamma^5 S_u (p+\frac{Q}{2}) \gamma^5 S_u (p+\frac{Q}{2}) \right] ~,
\nonumber
\eea
\noindent
after the explicit calculation of the trace in the last equation, one obtains
\bea
f_\pi = -4M_u g_{\pi\bar{q}q}(m_\pi^2) I_2^{uu}(m_\pi^2)
\label{fpi_eq}
\eea
where the coupling strength for the meson-quark-quark interaction is given by
\bea
g_{\pi\bar{q}q}^{-2}(m_\pi^2) = \frac{\partial \Pi_{\pi}^p}{\partial Q^2} \Big\vert_{Q^2=m_\pi^2}
= 2 \frac{\partial  \Pi_{uu}^p}{\partial Q^2}  \Big\vert_{Q^2=m_\pi^2} ~.
\eea
The last expression follows from the derivative of Eq.~\eqref{Pi_ij}:
\bea
\frac{\partial  \Pi_{uu}^p}{\partial Q^2}  = - 2 I_2^{uu} - 2 Q^2 \frac{\partial}{\partial Q^2} I_2^{uu}~.
\nonumber 
\eea

\subsection{$\eta$ and $\eta'$ mesons}

For the case of $\eta$ and $\eta'$ meson, the inverse mesonic propagator assumes a 
matrix form\cite{reviews} 
\bea
D^{-1} = \frac{1}{2} K^{-1} - \Pi ~,
\eea
  \noindent
where the effective coupling matrix $K$ and the mesonic self energy $\Pi$ for the $\eta-\eta'$ system are 
given by the matrices:
\bea
\Pi &=& \begin{pmatrix} \Pi_{00} & \Pi_{08} \\ \Pi_{08} & \Pi_{88} \end{pmatrix}~~,~~ 
K = \begin{pmatrix} K_{00} & K_{08} \\ K_{08} & K_{88}, \end{pmatrix} \nonumber ~,\\
K^{-1} & = &\frac{1}{\det K}\begin{pmatrix} K_{88} & -K_{08} \\ -K_{08} & K_{00} \end{pmatrix}
\eea 
where the determinant of $K$ is given by $\det K = K_{00}K_{88} - K_{08}^2$ and the  specific 
components are given by 
\bea
K_{00} &=& G + \frac{1}{3}K (2\sigma_u + \sigma_s),\\
K_{88} &=& G - \frac{1}{6}K (4\sigma_u - \sigma_s),\\
K_{08} &=& K_{80} =  -\frac{\sqrt{2}}{6}K (\sigma_u - \sigma_s),
\eea
and 
\bea
\Pi_{00} &=& \frac{2}{3}\left[2\Pi_{uu}^p(Q) +\Pi_{ss}^p(Q)\right],\\
\Pi_{88} &=&  \frac{2}{3}\left[\Pi_{uu}^p(Q) +2\Pi_{ss}^p(Q)\right],\\
\Pi_{08} &=& \Pi_{80} = \frac{2\sqrt{2}}{3}\left[\Pi_{uu}^p(Q) -\Pi_{ss}^p(Q)\right].
\eea
So, finally the inverse propagator is given by 
\bea
D^{-1} = \frac{1}{2\det K}\begin{pmatrix} A & B \\ B & C \end{pmatrix}    ~,
\eea 
with
\bea
A &=& K_{88} - 2 \det K ~ \Pi_{00},\\
B &=& -( K_{08} + 2 \det K ~\Pi_{08}),\\
C &=& K_{00} - 2 \det K ~\Pi_{88},
\eea
Diagonalizing $D^{-1}$ we get 
\bea
D^{-1} &=& \frac{1}{2\det K}\mathcal{O}^{-1}\begin{pmatrix} D_\eta^{-1} & 0 \\ 0 & D_{\eta'}^{-1} 
\end{pmatrix}\mathcal{O}
\eea 
where the diagonalizing orthogonal matrix $\mathcal{O}$ is given by 
\bea
\mathcal{O} &=& \begin{pmatrix} \cos\theta_p & \sin\theta_p \\ -\sin\theta_p & \cos\theta_p \end{pmatrix} ~,~
\tan (2\theta_p) = \frac{2B}{A-C} \nonumber
\eea 
and the diagonal elements now represent the dispersion relations for $\eta$ and $\eta'$ mesons, i.e. 
\bea
D_{\eta}^{-1} &=& (A+C) - \sqrt{(C-A)^2+4B^2},\\
D_{\eta'}^{-1} &=& (A+C) + \sqrt{(C-A)^2+4B^2}.\label{disp_metap_1}
\eea
The masses of the $\eta$ and $\eta'$ meson can now be determined from the equations 
\bea
D_{\eta}^{-1}(Q=m_\eta) &=& 0,\label{disp_meta}\\
D_{\eta'}^{-1}(Q=m_{\eta'}) &=& 0.\label{disp_metap}
\eea

At this point we want to note that $m_\eta$ can be evaluated from Eq.~\eqref{disp_meta} by using the 
integrals from Eqs.~\eqref{eqI1} and~\eqref{eqI2}. But as $\eta'$, in general, exists above the $\bar{q}q$
continuum, Eq.~\eqref{disp_metap} has complex poles, which we can assume to be of the form, 
$Q=Q_0 = m_{\eta'}-\frac{1}{2}i\Gamma$, with $\Gamma$ being the width of the $\eta'$-resonance.  
In the latter case the calculation of $A,B$ and $C$ in Eq.~\eqref{disp_metap_1} can be readily done
making  in Eq.~\eqref{Pi_ij}  the replacement:
\bea
Q^2 I_2^{ii} &\rightarrow& \left[m_{\eta'}^2 \textrm{Re}I_2^{ii} +
m_{\eta'} \Gamma \textrm{Im}I_2^{ii} \right]   \\
&+&
i\left[m_{\eta'}^2 \textrm{Im}I_2^{ii} - m_{\eta'}\Gamma \textrm{Re}I_2^{ii} \right] ~,
\nonumber
\eea
where terms of order $\Gamma^2$ have been neglected.
Then, after  substituting the latter expression in Eq.~\eqref{Pi_ij} one obtains
\bea
\Pi_{ii}^p(Q^2) = Re\Pi_{ii}^p(Q^2) + i Im\Pi_{ii}^p(Q^2) \nonumber ~,
\eea
where 
\bea
Re\Pi_{ii}^p(Q^2) &=& 4 I_1^i - 2 \left( m_{\eta'}^2 \textrm{Re}I_2^{ii} + 
m_{\eta'}\Gamma \textrm{Im}I_2^{ii} \right) \nonumber ~, \\
Im\Pi_{ii}^p(Q^2) &=& 2 \left(m_{\eta'}\Gamma \textrm{Re}I_2^{ii} - m_{\eta'}^2 \textrm{Im}I_2^{ii}\right) 
 ~.
\eea
The real and imaginary part of the integral $I_2^{ii}$  can be obtained from Eq.~\eqref{eqI2}   
by using the  Sokhotski-Plemelj formula 
\bea
\textrm{lim}_{\epsilon\rightarrow 0} \frac{1}{x-i\epsilon} = \mathcal{P}\frac{1}{x} + i\pi\delta(x),
\eea
where $\mathcal{P}$ stands for the Cauchy principal value and this formula makes sense only when integrated in $x$.
Then, one  obtains: 
\bea
Re I_2^{ii}(Q) &=& \frac{N_c}{2\pi^2}\mathcal{P}\int\limits_0^\Lambda \frac{p^2dp}{E_i}\frac{1}{Q^2-4E_i^2} 
\nonumber \\
Im I_2^{ii}(Q) &=&  \frac{N_c}{16\pi}\sqrt{1-\frac{4M_i^2}{Q^2}}   ~, \nonumber
\eea
where in these expressions we assume that  $Q$=$m_{\eta'}$ and $m_{\eta'}$ $>$ 2$M_i$.

\end{document}